\documentclass[twocolumn]{aastex631}
\usepackage{savesym}

\savesymbol{splitbox}
\usepackage{adjustbox}
\restoresymbol{SB}{splitbox}

\savesymbol{tablenum}
\usepackage{siunitx}
\restoresymbol{SIX}{tablenum}

\begin{document}

\title{The Size-Mass relation at Rest-Frame $1.5\mu$m from JWST/NIRCam in the COSMOS-WEB and PRIMER-COSMOS fields}

\author[0000-0003-2373-0404]{Marco Martorano}
\affiliation{Sterrenkundig Observatorium, Universiteit Gent, Krijgslaan 281 S9, 9000 Gent, Belgium}
\author[0000-0002-5027-0135]{Arjen van der Wel}
\affiliation{Sterrenkundig Observatorium, Universiteit Gent, Krijgslaan 281 S9, 9000 Gent, Belgium}
\author[0000-0002-3930-2757]{Maarten Baes}
\affiliation{Sterrenkundig Observatorium, Universiteit Gent, Krijgslaan 281 S9, 9000 Gent, Belgium}
\author[0000-0002-5564-9873]{Eric F. Bell}
\affiliation{Department of Astronomy, University of Michigan, 1085 South University Avenue, Ann Arbor, MI 48109-1107, USA}
\author[0000-0003-2680-005X]{Gabriel Brammer}
\affiliation{Cosmic Dawn Center (DAWN), Niels Bohr Institute, University of Copenhagen, Jagtvej 128, Kø benhavn N, DK-2200, Denmark}
\author[0000-0002-8871-3026]{Marijn Franx}
\affiliation{Leiden Observatory, Leiden University, P.O. Box 9513, 2300 RA, Leiden, The Netherlands}
\author[0000-0001-6843-409X]{Angelos Nersesian}
\affiliation{STAR Institute, Universit\'e de Li{\`e}ge, Quartier Agora, All\'ee du six Aout 19c, B-4000 Liege, Belgium}
\affiliation{Sterrenkundig Observatorium, Universiteit Gent, Krijgslaan 281 S9, 9000 Gent, Belgium}

\correspondingauthor{Marco Martorano}
\email{marco.martorano@ugent.be}

\begin{abstract}
We present the galaxy stellar mass - size relation in the rest-frame near-IR ($1.5~\mu{\text{m}}$) and its evolution with redshift up to $z=2.5$. S\'ersic profiles are measured for {$\sim$}$26\,000$ galaxies with stellar masses $M_\star > 10^9~{\text{M}}_\odot$ from JWST/NIRCam F277W and F444W imaging provided by the COSMOS-WEB and PRIMER surveys, using coordinates, redshifts, colors and stellar mass estimates from the COSMOS2020 catalog. 
The new rest-frame near-IR effective radii are generally smaller than previously measured rest-frame optical sizes, on average by 0.14~dex, with no significant dependence on redshift.
For quiescent galaxies this size offset does not depend on stellar mass, but for star-forming galaxies the offset increases from -0.1~dex at $M_\star = 10^{9.5}~{\text{M}}_\odot$ to -0.25~dex at $M_\star > 10^{11}~{\text{M}}_\odot$. That is, we find that the near-IR stellar mass - size relation for star-forming galaxies is flatter in the rest-frame near-IR than in the rest-frame optical at all redshifts $0.5<z<2.5$.
The general pace of size evolution is the same in the near-IR as previously demonstrated in the optical, with slower evolution ($R_{\text{e}} \propto (1+z)^{-0.7}$) for $L^*$~star-forming galaxies and faster evolution ($R_{\text{e}} \propto (1+z)^{-1.3}$) for $L^*$~quiescent galaxies. Massive ($M_\star>10^{11}~{\text{M}}_\odot$) star-forming galaxies evolve in size almost as fast as quiescent galaxies. Low-mass ($M_\star<10^{10}~{\text{M}}_\odot$)~quiescent galaxies evolve as slow as star-forming galaxies. Our main conclusion is that the size evolution narrative as it has emerged over the past two decades does not radically change when accessing the rest-frame near-IR with JWST, a better proxy of the underlying stellar mass distribution.

\end{abstract}


\section{INTRODUCTION} \label{sec:intro}

The size of a galaxy retains information about its evolutionary history and is partially a reflection of the properties of the host dark matter halo, most notably the virial radius and the angular momentum \citep{fall80, dalcanton97, mo98}. Galaxy sizes are therefore a key ingredient of the various well-known scaling relations such as the size-stellar mass relation \citep[e.g.][]{shen03, van-der-wel14, whitaker17, mowla19a, dimauro19, Nedkova21, magnelli23, shen23, ward23, cutler23} and the fundamental plane \citep{kormendy77, djorgovski87, dressler87}.

Parametric fits can give estimates of galaxies' size at the expense of having some sensitivity to the parametric form but with the advantage of being less sensitive to point-spread function (PSF) smearing and noise effects than nonparametric methods \cite[for a complete review, see][]{conselice14}. S\'ersic fits are the most common profiles adopted to fit galaxies. These are parameterized by flux, half-light radius ($R_{\text{e}}$), and S\'ersic index ($n$), which quantifies the concentration of the light profile.

While the S\'ersic index remains relatively constant when moving from optical wavelengths to the near-IR \citep{kelvin12, kennedy15, martorano23, yao23}, the half-light radius is significantly influenced by color gradients due to stellar population variations and attenuation by dust \citep{sandage72, peletier90,de-jong96a}.
Numerous low-redshift studies have analyzed the wavelength-dependent behavior of $R_{\text{e}} $ \citep[e.g.,][]{kelvin12, haussler13, pastrav13, vika13, vulcani14, kennedy15, lange15, baes20, nersesian23}. The general result is that galaxy sizes are smaller at longer wavelengths, regardless of type or mass. The implied color gradients are varyingly attributed to gradients in age, metallicity, star-formation activity, and dust attenuation, depending on the type and mass of the galaxy. 

Spatially resolved imaging is essential to study structural evolution at large lookback time to provide stringent constraints on galaxy evolution models. Only the Hubble Space Telescope (HST) and the James Webb Space Telescope (JWST) have sufficient spatial resolution ($\approx 1~{\text{kpc}}$) and stability of the PSF to accurately quantify the light profiles of such distant galaxies. HST has been extensively used over the past two decades to map the size evolution at rest-frame UV and optical wavelengths \citep{daddi05, trujillo06b, zirm07, van-dokkum08, van-der-wel08b, newman12, van-der-wel14, shibuya15, mowla19a, mowla19b, Nedkova21, cutler22}, where attenuation by dust and the outshining by young populations present significant obstacles for interpreting the observed sizes.
Now that JWST provides us with rest-frame near-IR observations of high-redshift galaxies, we can estimate sizes at longer wavelengths where dust attenuation is much less severe and the outshining effect lessened, such that gradients in the dust content and stellar population properties are suppressed.

The first estimates of rest-frame near-IR galaxy sizes from JWST by \cite{suess22} found a systematic 9\% reduction in the sizes at 4.4$\mu{\text{m}}$ compared to those at 1.5$~\mu{\text{m}}$ (both in the observed frame) in the redshift range $z=1-2.5$. The general result that galaxies are smaller in the rest-frame near-IR compared to the rest-frame optical was subsequently confirmed \citep{ito24, ormerod24, cutler23}, displaying the same behavior as for present-day galaxies \citep{evans94, mollenhoff06, kelvin14}.

In this work, we present rest-frame 1.5$~\mu{\text{m}}$ effective radius measurements for {$\sim$}$26\,000$ galaxies with stellar masses $M_\star\geq10^{9}~{\text{M}}_\odot$ in the redshift range $z=0.5-2.5$ that fall in the COSMOS-WEB \citep{casey22, casey23} and PRIMER-COSMOS \citep{dunlop21} JWST footprints. We take advantage of these new measurements to determine the rest-frame near-IR size-mass relation and its evolution with redshift for the general galaxy population, allowing for a direct comparison with the well-established results obtained in the rest-frame optical wavelength regime.

The paper is structured as follows. In Section \ref{sec:data} we briefly present the cataloged data used in this work along with an overview of the JWST observations. Section \ref{sec:JWSTfit} follows with an explanation of the fitting technique adopted to retrieve galaxies' sizes. In Section \ref{sec:results} we present the size-mass relation and the redshift evolution of the median size of galaxies. This is followed by a discussion in Section \ref{sec:discussion}. Finally, in Section \ref{sec:conclusion}, we summarize our findings and draw our conclusions. We adopt a flat $\Lambda$CDM cosmology with H$_0=70~$km$~$s$^{-1}~$Mpc$^{-1}$, $\Omega_m=0.3$.

\section{DATA AND SAMPLE SELECTION}\label{sec:data}

In this section, we describe the parent catalog that we used to select target galaxies and their basic properties (redshifts, stellar masses, star-formation activity). We also describe the JWST data used to fit near-IR light profiles.

\subsection{COSMOS2020} \label{sec:cosmos2020}
The parent sample is drawn from the \textit{COSMOS2020-CLASSIC} catalog presented by \cite{weaver22}.
It contains multiwavelength ($0.15~\mu m$ to $\sim8~\mu m$) observations of over 1.7 million sources in the 2~deg$^2$ COSMOS field \citep{scoville07, koekemoer07} from several ground-based (i.e. Subaru/HSC and VISTA) and space-based  (i.e. HST and Spitzer) observatories. The catalog includes stellar population parameters obtained from broadband photometry spectral energy distribution (SED) fitting with the codes \textsc{LePhare} \citep{arnouts02, ilbert06} and \textsc{Eazy} \citep{brammer08}. Here we use the redshifts and stellar population parameters from \textsc{LePhare} as those are preferred by the authors \citep{weaver22}. \textsc{Eazy} provides the corresponding rest-frame colors U$-$V and V$-$J. {Mixing stellar masses from \textsc{LePhare} with rest-frame colors from \textsc{Eazy} does not bias our results: the \textsc{Eazy} stellar mass estimates produce the same conclusions, but \citet{weaver22} prefer \textsc{LePhare} because of its lower redshift bias fraction. Furthermore, the \textsc{LePhare} stellar mass estimates are closer to those used by \cite{van-der-wel14}, which we use as a comparison sample in this work (Appendix~\ref{app-sec: Fast-LePhare}).}

\subsection{JWST observations}\label{sec:JWSTphoto}
\begin{figure}
    \centering
    \includegraphics[scale=0.5]{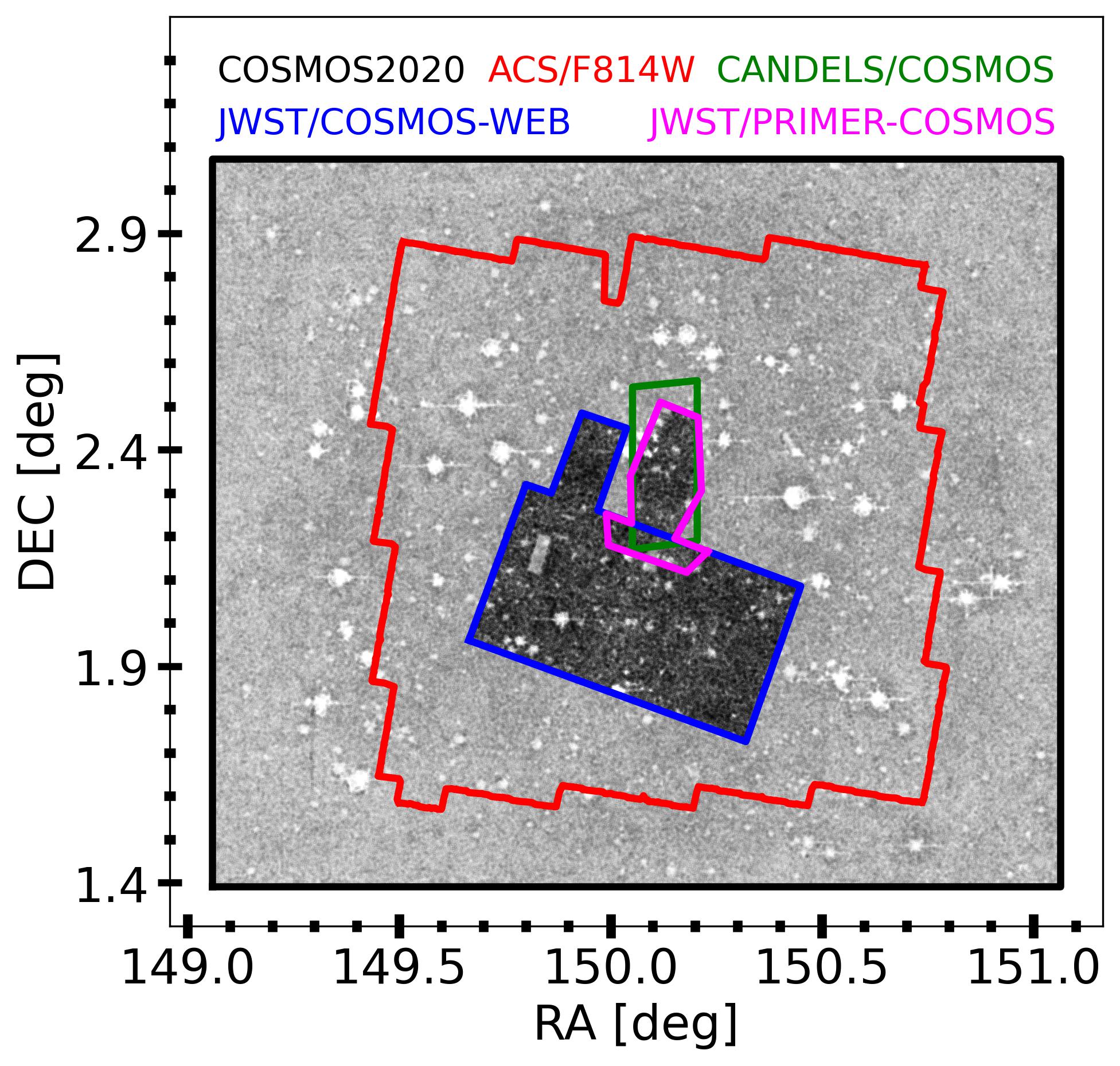}
    \caption{Overview of the footprints of the surveys used in this work. The black square shows the full COSMOS2020 field; the red contours represent the COSMOS2020 field observed with HST-ACS/F814W. The green area identifies the CANDELS-COSMOS field. Blue and magenta lines are used to identify, respectively, the JWST footprints of the COSMOS-WEB and PRIMER-COSMOS programs. Gray dots in the background show the positions of galaxies identified in the COSMOS2020 catalog while darker dots highlight the position of galaxies observed in the long-wavelength channel of JWST/NIRCam.}
    \label{fig:footprint}
\end{figure}
A region of $\approx0.54$~deg$^2$~roughly in the middle of the $\approx 2$~deg$^2$~COSMOS field (Fig.~\ref{fig:footprint}) is covered by JWST/NIRCam imaging from the COSMOS-WEB program \citep[GO:1727,][]{casey23}. COSMOS-WEB provides imaging with F115W and F150W in the short-wavelength channel and with F277W and F444W in the long-wavelength channel.
The 5$\sigma$ depth reached is between 27.61 and 28.18 ABmag for the filter F444W depending on the number of exposures \citep{casey23}. Currently, slightly more than half of the $\approx0.54$~deg$^2$ area has been observed and will be used in this work.

The CANDELS-COSMOS field \citep{grogin11, koekemoer11}, consisting of a deeper but smaller field within COSMOS, was observed by JWST as part of the PRIMER program \citep[GO:1827,][]{dunlop21} with 10 different JWST/NIRCam filters and an {F444W depth up to 28.17}~ABmag \citep{cutler23} for the regions with the highest exposures. 

Data from these two programs were processed with \textsc{grizli} \citep{grizli} to create mosaics and weight maps with a pixel scale of \SI{0.05}{\arcsecond}/pix.
The mosaics are publicly available on \textit{The DAWN JWST Archive} \citep[DJA][]{valentino23}. For this work, we retrieve cutouts for each galaxy from the \textsc{Interactive Map Interface} \citep{hausen22} available on the \textit{DJA}. 
Focusing on the rest-frame near-IR, in this paper, we only use the long-wavelength channel filters F277W and F444W. As described below, noise properties and PSF are tested and well understood for this dataset.

\begin{figure*}
    \centering
    \includegraphics[scale=0.24]{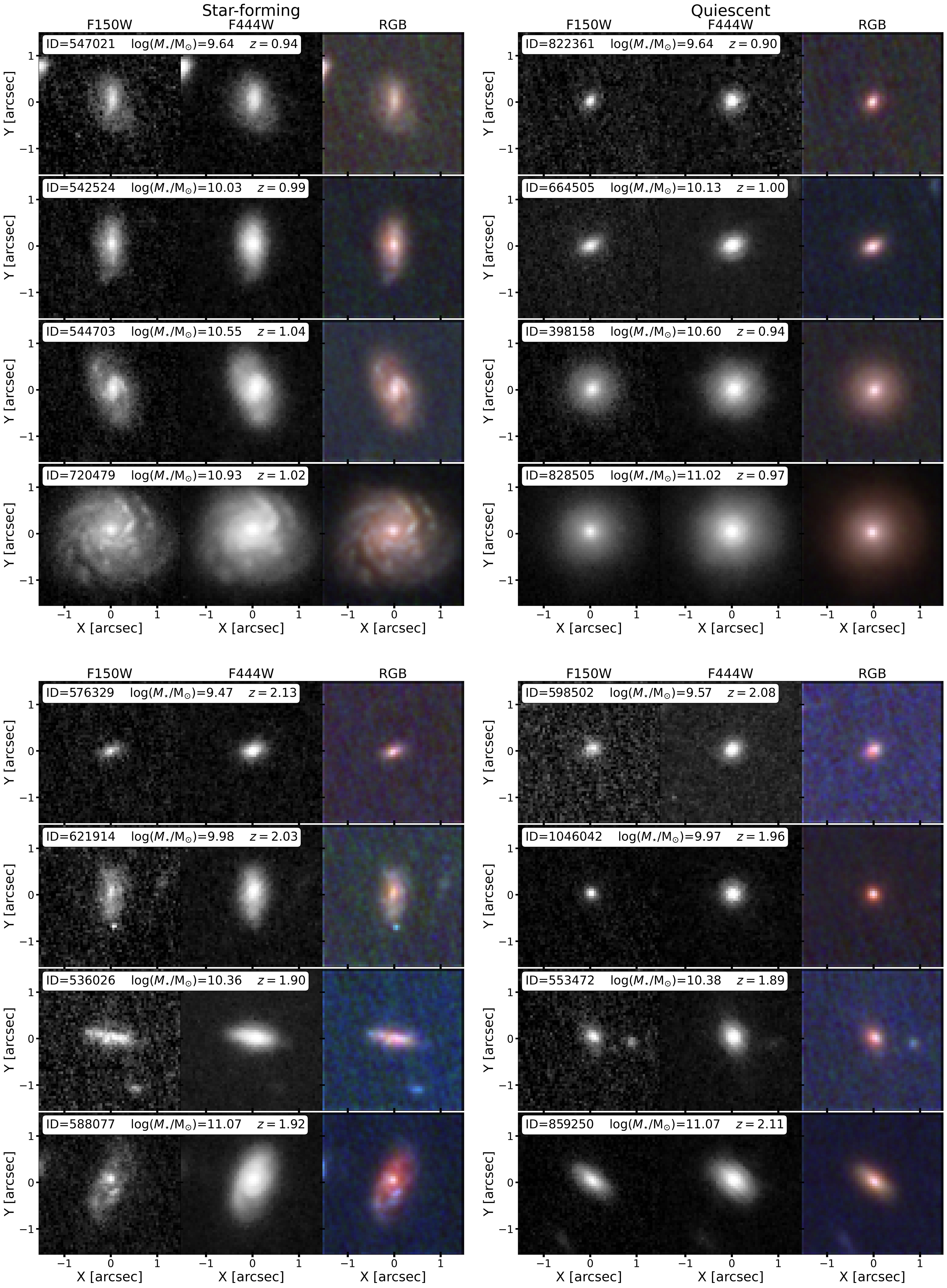}\caption{JWST/NIRCam imaging examples for 16 galaxies selected to be representative of the median characteristics of the sample. The top four rows show galaxies at $z\sim1$ while the bottom four rows show galaxies at $z\sim2$. For each galaxy, the COSMOS2020 ID, \textsc{LePhare} stellar mass and redshift are indicated. Filters F150W and F444W are shown as representatives of the optical and near-IR wavelengths. The RGB frame is obtained by combining the filters F277W, F150W, and F115W with a Richardoson-Lucy deconvolution.
    }
    \label{fig:rgb_imgs}
\end{figure*}

Figure \ref{fig:rgb_imgs} shows JWST/NIRCam imaging for 16 galaxies in the sample selected to be representative of the median characteristics of galaxies across the full mass and redshift range (see Sec.~\ref{sec:discussion} for further discussion). JWST/NIRCam short-wavelength channel filters F115W and F150W are used only for visualization and are not used for quantitative analysis in this work.

\subsection{Galaxy selection} \label{sec:gal_sel}

\begin{figure}
    \centering
    \includegraphics[scale=0.5]{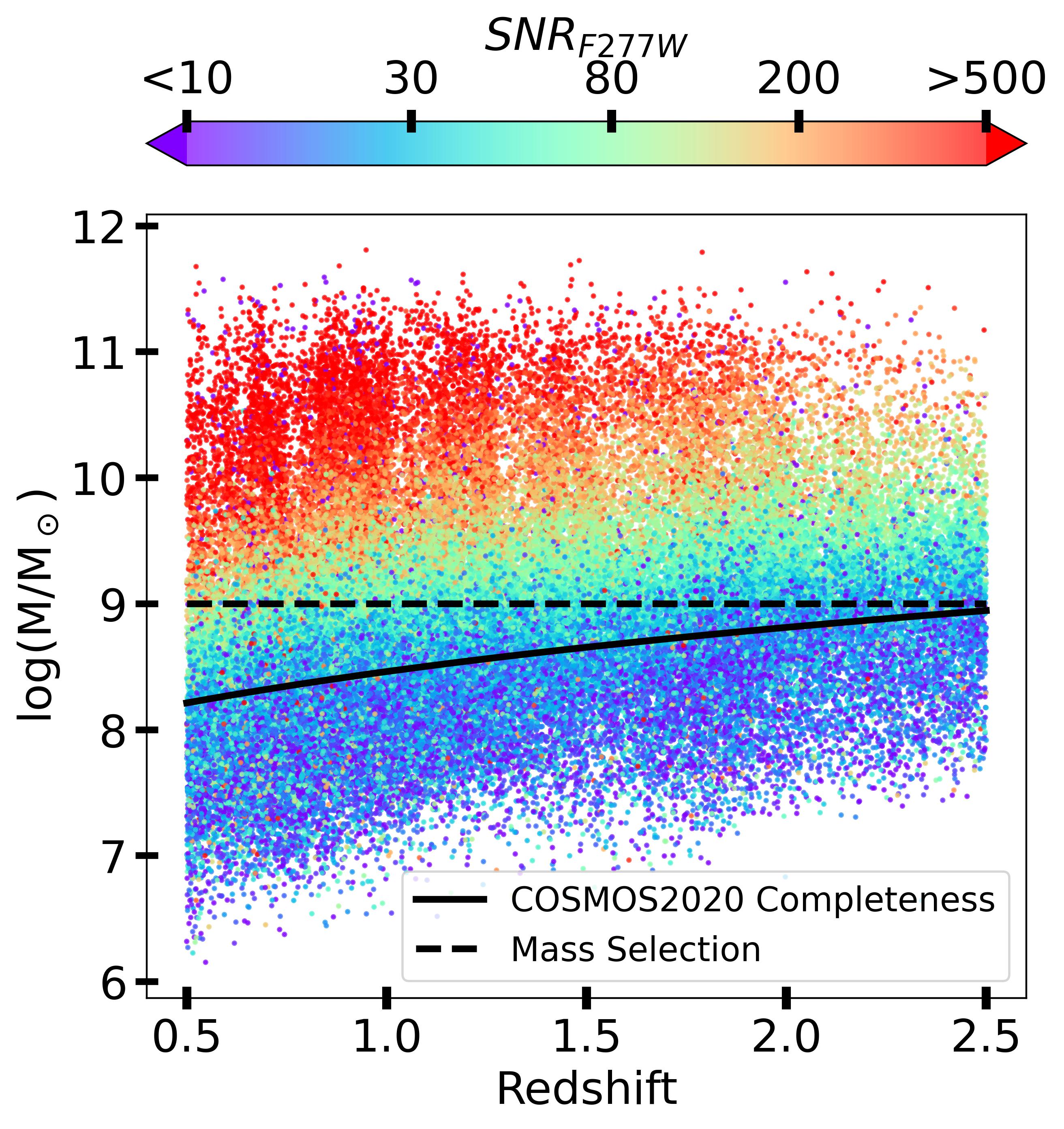}
    \caption{Stellar mass-redshift distribution of the sample of galaxies for which \textsc{GalfitM} light profiles were fit, color-coded with the SNR measured from the JWST/NIRCam F277W image. The solid black line shows the stellar mass completeness limit from \cite{weaver22}. The dashed line shows the lower-mass limit adopted in this paper.}
    \label{fig:mass_sel}
\end{figure}

Since this paper aims to present near-IR sizes at rest-frame 1.5$\mu{\text{m}}$ the wavelength coverage of the F277W and F444W filters imposes a redshift range of $z=0.5-2.5$ to avoid the effects of extrapolation. 
The rest-frame wavelength corresponding to the pivot wavelength of F277W is 1.85$~\mu{\text{m}}$ at $z=0.5$ and that of F444W 1.26$~\mu{\text{m}}$ for $z=2.5$.

Among the over 1.7 million sources in the \textit{COSMOS2020-CLASSIC} catalog, we select the $75\,765$ galaxies (ACS$\_$MU$\_$CLASS=1{; this flag rejects stellar-like objects)} in this redshift range that have a positive HST/F814W flux and fall within the available JWST/NIRCam F277W and F444W imaging footprints (Fig. \ref{fig:footprint}).

In Figure \ref{fig:mass_sel} we show the mass-redshift distribution of these galaxies. The adopted stellar mass limit for our sample ($M_\star~=~10^{9}~{\text{M}}_\odot$) leaves us with $32\,002$ galaxies.

We remove from our sample 4912 objects ($\sim$15$\%$) due to differences between the identification, deblending, and segmentation between the source catalog from \cite{weaver22} and our work based on the much-higher-resolution NIRCam imaging (see \ref{sec:cutouts}). This rejection does not bias our sample, since these targets are evenly distributed across the full mass and redshift range. Moreover, mismatches are often related to nearby bright sources and/or severe blending that prevents robust profile fits in the first place. NIRCam-selected catalogs will eventually produce a clear improvement in object selection and characterization, but this is beyond the scope of this work.

Size estimates may be biased due to the contributions from central point sources from active galactic nuclei (AGN). Starlike objects were already removed from the sample, so that the most obvious AGN are automatically omitted. But AGN contributions in the mid-IR still have to be addressed since \cite {chang17} showed that an AGN fraction $>50\%$ induces a radius decrease up to $50\%$. We match our sample to the publicly available catalog presented in \cite{chang17}, who fit \textsc{MAGPHYS} on the COSMOS2015 \citep{laigle16} photometry for $36\,000$ sources in the COSMOS field. We find, and remove from our sample, 45 galaxies with the median of the posterior distribution of the AGN fraction larger than $50\%$. Since these 45 galaxies show no obvious bias in the size distribution, we conclude that more moderate AGN contributions will not affect our results either.

\subsection{Separating Star-Forming and Quiescent Galaxies}\label{sec:gal_class}

\begin{figure*}
    \centering
    \includegraphics[scale=0.38]{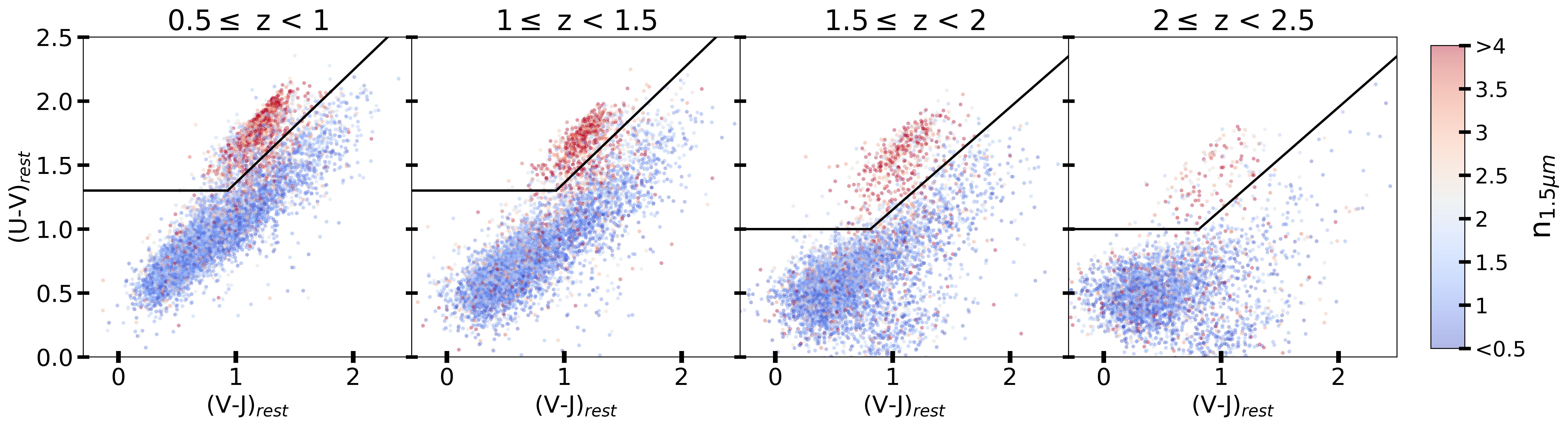}
    \caption{$UVJ$ diagram for our selected sample of galaxies color-coded with the S\'ersic index measured at rest-frame 1.5$~\mu{\text{m}}$. Solid lines identify the quiescent locus in the $UVJ$ plane as described in the text.} \label{fig:UVJ}
\end{figure*}

The COSMOS2020 SED fits from \cite{weaver22} do not include mid- or far-IR photometric data; consequently, the resulting star formation rates are not robust enough to be efficiently used to divide galaxies in star-forming and quiescent. Instead, we separate star-forming and quiescent galaxies based on their rest-frame $U-V$ and $V-J$ colors.

\cite{williams09} defined selection criteria to separate the quiescent and the star-forming population from the $UVJ$ diagram. \cite{muzzin13} adapted the exact constraints to better match their rest-frame photometry. But we find that those cuts do not work well for the current rest-frame photometry. {This difference originates from two main factors: the COSMOS2020 photometry is different from the one used in \cite{muzzin13} despite the similar sample (different methodology for photometry retrieval, several new filters, updated zero-points), and the templates used for SED fitting in COSMOS2020 are different from those used by \cite{muzzin13} leading to different rest-frame colors. These two effects combined necessitate new color cuts to maximize the separation between the two galaxy types. Note that drawing these separation lines has always been a heuristic process.}
In Figure \ref{fig:UVJ} we present the $UVJ$ diagram and the new color cuts adopted to optimize the separation between the two types. 
The quiescent region is finally identified as
\begin{equation}\label{eq:UVJ_lowz}
    z<1.5:    \;\;\;  U-V > 1.3; \;\; U-V>0.48+0.88(V-J) 
\end{equation}
\begin{equation}\label{eq:UVJ_highz}
    z\geq1.5: \;\;\; U-V > 1; \;\, \;\;\; U-V>0.35+0.80(V-J)
\end{equation}

{The original \cite{muzzin13} color classification only leads substantially different sample separation at $z>2$, due to the many quiescent galaxies with relatively blue colors.}

\section{Light Profile Fits} \label{sec:JWSTfit}
In this section, we present the steps of the light fitting procedure wit\textsc{GalfitM} \citep{haussler13, vika13}, a further developed version of \textsc{Galfit} \citep{peng02,peng10}.

\subsection{Cutouts}\label{sec:cutouts}

For each galaxy, we create F277W and F444W cutouts such that the number of pixels not associated with a source in the F444W segmentation map \citep[created with SEP, assuming a detection threshold of 3$\sigma$, a minimum area of 5~pixels and a minimum contrast ratio used for object deblending of 0.02;][]{bertin96, barbary16} is at least 100 times the number of pixels that are associated with object segments. In addition, we impose a minimum size of \SI{8}{\arcsecond} and a maximum size of \SI{20}{\arcsecond}.
A posteriori, we verify that 99\%(50\%) of the cutouts are at least 12(37) times larger than $R_{\text{e}}$, which guarantees that uncertainties in background estimates propagate into uncertainties in $R_{\rm{e}}$ by less than 5\%($<$1\%) \citep[see Fig. 6 in][]{george24}.
The 1\% of galaxies with cutouts smaller than 12 times $R_{\text{e}}$ are large ($R_{\text{e}}>5~{\text{kpc}}$) low-mass galaxies ($M_\star<10^{10}~{\text{M}}_\odot$) mostly at $z<1.5$. A systematic $5\%$ variation in the size of these targets has a negligible effect ($<0.5\%$) on the median stellar mass-size relation presented in \S\ref{sec:size-mass}.

For each cutout, we use the available weight (inverse variance) image to compute an error map. 
Masks are created based on the segmentation map, including those sources that are more than a magnitude fainter than the target, as also done in \cite{martorano23}.
Sources brighter than this threshold are included in the \textsc{GalfitM} profile fit as separate objects, along with the target itself, and fitted simultaneously. 
\textsc{SEP} provides us with an initial estimate for \textsc{GalfitM} of the semi-major and minor axes of the galaxies as well as their center coordinates.

\begin{figure}
    \centering
    \includegraphics[scale=0.38]{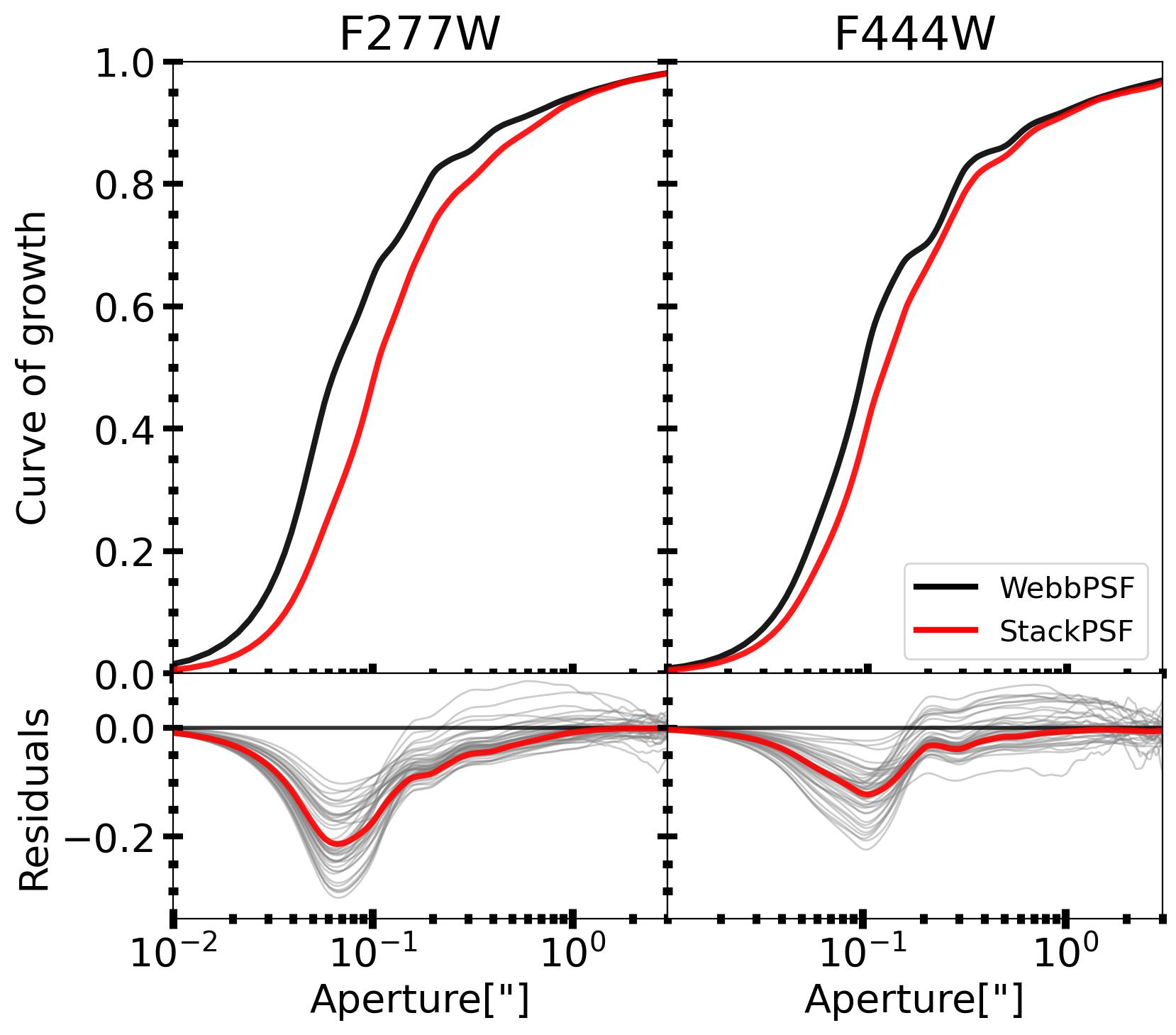}
    \caption{Curves of growth for the theoretical PSF profile (WebbPSF; solid black lines) and our stacked PSF (StackPSF; solid red line) constructed from 50 stellar light profiles (light grey lines) in the two NIRCam filters F277W (left panel) and F444W (right panel). In the lower panels, we show the residuals of the curve of growth of our PSF with respect to the theoretical one. The empirical PSFs are significantly broader than the theoretical PSFs.}
    \label{fig:PSF}
\end{figure}

We find 144 galaxies to have at least 1 pixel with flux zero within a square of side 7 pixels from the center recovered with \textsc{SEP}. These pixels can either be bad pixels masked by the pipeline used to create the mosaics or gaps in the footprint. Since the effect on the fit produced by a masked pixel in the inner part of the galaxy can be large, we decided to remove these 144 sources from our sample.

\begin{figure*}
    \centering
    \includegraphics[scale=0.48]{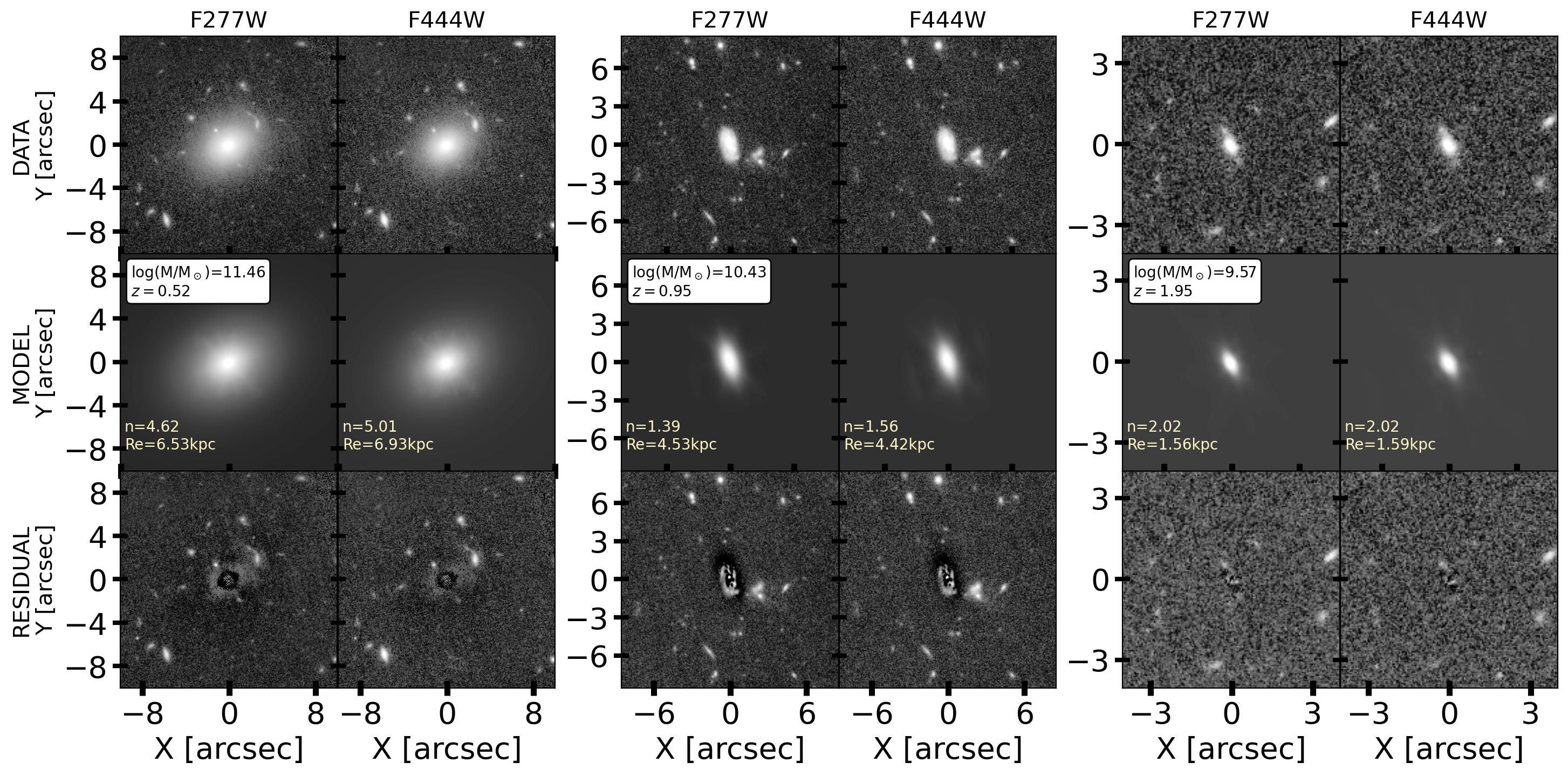}
    \caption{\emph{Left two columns:} Galaxy 472142 (from COSMOS2020) in JWST/NIRCam filters F277W and F444W. We show, from top to bottom, the image , the 2D light model from \textsc{GalfitM} and the residuals of the fit. \emph{Middle columns:} COSMOS2020 ID 615747. \emph{Right columns:} COSMOS2020 ID 515598. Cutouts are presented in the same size used for the morphological fit performed with \textsc{GalfitM}.}
    \label{fig:3gal}
\end{figure*}

\subsection{The Point Spread Function}

\textsc{GalfitM} generates S\'ersic profiles that are convolved with the PSF of the filter investigated and subsequently fitted to the image. 
As several works pointed out \citep{nardiello22,weaver23,zhuang23,ji23}, we also find the PSF in the mosaics to be broader than the theoretical PSFs from \textsc{WebbPSF} \citep{perrin14}, which is primarily the result of the data reduction process (shifting, weighting, and stacking).
To minimize the systematic effects that a wrongly calibrated PSF can induce, we create empirical PSFs. We visually select 50 stars across the mosaic that are not saturated and that overlap flux density distribution with the sample galaxies, as the PSF might be flux-dependent. Subpixel precisions of the centers of the stars are determined by \textsc{SEP} run on \SI{8}{\arcsecond} cutouts.
The flux is normalized in each cutout to the flux measured within a \SI{0.3}{\arcsecond} circle around the \textsc{SEP} center. After masking any other source in the cutout, we compute the background as the median flux per pixel in an annulus with inner radius \SI{2.5}{\arcsecond} and outer radius \SI{3}{\arcsecond}. This median background is subtracted from all the pixels. Each cutout is then upsampled by a factor 10 and cropped to a square of \SI{6}{\arcsecond} centered with subpixel resolution. We then produce a median stack (on a pixel-by-pixel basis) of the 50 upsampled, background-subtracted cutouts. Our final empirical PSF is then produced by a normalization such that the total flux in the \SI{6}{\arcsecond} square matches that of the theoretical PSF (generated via \textsc{WebbPSF}) in the same area. {Exposures from COSMOS-WEB and PRIMER-COSMOS have the same pointing position angle (differences are $<1^{\circ}$), so that the stacked stars are well aligned.}

In Figure \ref{fig:PSF} we compare our empirical PSF with the theoretical \textsc{WebbPSF} PSF. The large differences in enclosed light (up to $\approx 20$\% at \SI{0.07}{\arcsecond} for F277W) imply that the \textsc{WebbPSF} PSF cannot be used in combination with these mosaics to infer accurate structural properties of galaxies. As referred to before, these differences can be attributed to the weighting and stacking procedure of multiple exposures in the data reduction pipeline and do not in any way reflect an error in the \textsc{WebbPSF} models.

The variation in stellar light profiles reflects, for one, spatial variation in the PSF. The median and 16-84 percentile ranges in the radius encircling half of the stellar light profile are $\SI{0.105\pm 0.016}{\arcsecond}(\SI{0.122 \pm 0.022}{\arcsecond})$. We interpret the scatter among the stellar profiles as the uncertainty in the PSF width  ($15\%$ and $18\%$ in F277W F444W). These variations are comparable to the 20\% spatial variation across the detector itself found by \cite{nardiello22}. The $\SI{0.016}{\arcsecond}$ and $\SI{0.022}{\arcsecond}$ are propagated into the galaxy size uncertainties, by adding in quadrature to the formal fitting uncertainties. The PSF uncertainty dominates the random uncertainty for small objects $\lesssim 1$~kpc, but does not affect the median trends that are the main result of this paper.

\subsection{Profile Fits with \textsc{GalfitM}}

The convolution box used by \textsc{GalfitM} is set to be 200 pixels per side. Despite slowing down the computation, the large convolution box was adopted to encompass even the largest sources. We can then use the same convolution box for all the targets, avoiding the introduction of systematics due to the use of different convolution boxes for different galaxies.

S\'ersic profile fits are independently fitted to the F277W and F444W images, allowing the S\'ersic index to span the range 0.2 - 10 with an initial guess of 2.3. The x and y coordinates of the galaxies' centers are limited to lie within a distance of 5~pixels from the \textsc{SEP}-based centers. The effective radius is limited to values between 0.01 and 200 pixels with the initial value defined as the semi-major axis of the \textsc{SEP} ellipse.
Axis ratio and position angle are left as free parameters with no constraints. The mosaics are background-subtracted, but to allow for remaining background patterns across the mosaic, we adopt a constant background with \textsc{GalfitM} as a free parameter.

We find that the fits {reach the numerical convergence} for both JWST/NIRCam filters for {$26\,003$} galaxies ($96\%$ of the available sample). 

We find, and remove, one galaxy with a retrieved S\'ersic index within $10^{-3}$ from the boundaries of the parameter space suggesting the sampler got stuck there.

For 514 sources, \textsc{GalfitM} produces a fit with {$\chi_{reduced}^2 >3$}. A visual inspection of these fits reveals that for the vast majority (506) the {$\chi_{reduced}^2$} is driven by those galaxies fitted simultaneously with the target and not by the target itself. We refit these sources, masking the nearby galaxies, and we find that they converge properly, with much lower {$\chi_{reduced}^2$} for 497 out of 506. The remainder are omitted from the final sample.

Visual inspection by M.M. of all {$26\,003$} fitted galaxies shows 36 mergers and two lensing systems that are not properly fitted and are, therefore, discarded (six mergers and the two lensing systems were among those fits with {$\chi_{reduced}^2>3$}). We further reject five galaxies that were only partially covered by the NIRCam mosaics, leaving the fitting results unreliable.
Our final sample with reliable S\'ersic profiles includes {$25\,960$} galaxies of which {$3\,082$} are quiescent and {$22\,878$} are star-forming.

In Figure \ref{fig:3gal} we show the \textsc{GalfitM} fitting results for three randomly selected galaxies. Although low-redshift high-mass galaxies are bright and extended, our pipeline uses sufficiently large cutouts to enable accurate background estimates. At the same time, for low-mass high-redshift galaxies, we see that the signal-to-noise ratio (SNR) is sufficiently high for a precise size estimate {(larger than 30 (50) for 84\% (50\%) of galaxies with $M_\star<9.5~{\text{M}}_\odot$ and redshift $z>2$)}.

The uncertainties retrieved for S\'ersic index ($n$) and effective radius are systematically underestimated by \textsc{GalfitM} \citep{van-der-wel12}. Following the prescriptions in \cite{van-der-wel12} and \cite{martorano23}, we reassess the uncertainties: we adopt a baseline uncertainty of 0.1~dex on the S\'ersic index for galaxies with SNR~=~50 and on $R_{\text{e}} $ for galaxies with SNR~=~20, and compute the final uncertainties as $\propto 1/\sqrt{\rm{SNR}}$. This methodology is based on an analysis of HST/WFC3 data \citep{van-der-wel12}. The SNR of JWST/NIRCam is much higher, producing, at times, very small uncertainties. In fact, for $\sim32\%(5\%)$ of the galaxies, the \textsc{GalfitM} uncertainty on $R_{\text{e}}$($n$) is larger than the SNR-based value, and we retain the former.  

The rest-frame 1.5$~\mu{\text{m}}$ $R_{\text{e}} $ and $n$ are calculated by fitting a straight line to the values of $R_{\text{e}} $ and $n$ in F277W and F444W using the respective pivot wavelengths. This fit is repeated 100 times, sampling from the F277W and F444W $R_{\text{e}} $ and $n$ uncertainties, adopting the median values of the $R_{\text{e}}$ and $n$ polynomials at 1.5$~\mu{\text{m}}$ as the estimates and half of the 16th-84th percentile range as the 1$\sigma$~uncertainty.

In Table \ref{tab:galfitmonline}  we present \textsc{GalfitM} outputs produced from the fitting of the NIRCam/F277W, F444W cutouts and the rest-frame $1.5\mu{\text{m}}$ effective radii and S\'ersic indices computed as discussed above. A full version of the table is available online in the machine-readable format.
\begin{deluxetable}{clp{0.66\linewidth}}
\tablewidth{0pt}
\tablecaption{\textsc{GalfitM} outputs produced from the fitting of the NIRCam/F277W, F444W cutouts, and the rest-frame $1.5\mu{\text{m}}$ effective radii and S\'ersic indices \label{tab:galfitmonline}}
\tablehead{
\colhead{Label} &
\colhead{Unit} & 
\colhead{Description}
}
\startdata
   Classic        & ---    &  COSMOS2020/CLASSIC Identifier [ID] \\ 
   RAdeg          & deg    &  COSMOS2020/CLASSIC Right Ascension, decimal degrees (J2000) [ALPHA\_J2000] \\ 
   DEdeg          & deg    &  COSMOS2020/CLASSIC Declination, decimal degrees (J2000) [DELTA\_2000] \\ 
   lpzBEST        & ---    &  COSMOS2020/CLASSIC LePhare photo-z redshift [lp\_zBEST] \\ 
   loglpMassbest  & [Msun] &  Corrected (see text) COSMOS2020/CLASSIC LePhare BC03 log stellar mass [lp\_mass\_best] \\ 
   Re-F277W       & arcsec &  GalfitM effective radius in filter F277W \\ 
e\_Re-F277W       & arcsec &  Uncertainty in Re-F277W \\ 
   n-F277W        & ---    &  GalfitM S\'ersic index in filter F277W \\ 
e\_n-F277W        & ---    &  Uncertainty in n-F277W \\ 
   mag-F277W      & mag    &  GalfitM AB magnitude in filter F277W \\ 
e\_mag-F277W      & mag    &  Uncertainty in mag-F277W \\ 
   q-F277W        & ---    &  GalfitM axis ratio in filter F277W \\ 
e\_q-F277W        & ---    &  Uncertainty in q-F277W \\ 
   PA-F277W       & deg    &  GalfitM position angle in filter F277W  \\ 
e\_PA-F277W       & deg    &  Uncertainty in PA-F277W \\ 
   Re-F444W       & arcsec &  GalfitM effective radius in filter F444W \\ 
e\_Re-F444W       & arcsec &  Uncertainty in Re-F444W \\ 
   n-F444W        & ---    &  GalfitM S\'ersic index in filter F444W \\ 
e\_n-F444W        & ---    &  Uncertainty in n-F444W \\ 
   mag-F444W      & mag    &  GalfitM AB magnitude in filter F444W \\ 
e\_mag-F444W      & mag    &  Uncertainty in mag-F444W \\ 
   q-F444W        & ---    &  GalfitM axis ratio in filter F444W \\ 
e\_q-F444W        & ---    &  Uncertainty in q-F444W \\ 
   PA-F444W       & deg    &  GalfitM position angle in filter F444W  \\ 
e\_PA-F444W       & deg    &  Uncertainty in PA-F444W \\ 
   Re-1.5um       & arcsec &  Effective radius at rest-frame $1.5\mu{\text{m}}$ \\ 
e\_Re-1.5um       & arcsec &  Uncertainty in Re-1.5um \\ 
   n-1.5um        & ---    &  S\'ersic index at rest-frame $1.5\mu{\text{m}}$ \\ 
e\_n-1.5um        & ---    &  Uncertainty in n-1.5um \\ 
\enddata
\tablecomments{\small{The 
original COSMOS2020/CLASSIC labels are given in square brackets in the
description of the columns taken or derived from that catalog. \\
Table \ref{tab:galfitmonline} is published in its entirety in 
the machine-readable format. A description of the columns is provided here 
for guidance regarding its form and content.}}
\end{deluxetable}

\section{RESULTS} \label{sec:results}
\subsection{Size-Mass Distribution}\label{sec:size-mass}

\begin{figure*}
    \centering
    \includegraphics[scale=0.30]
    {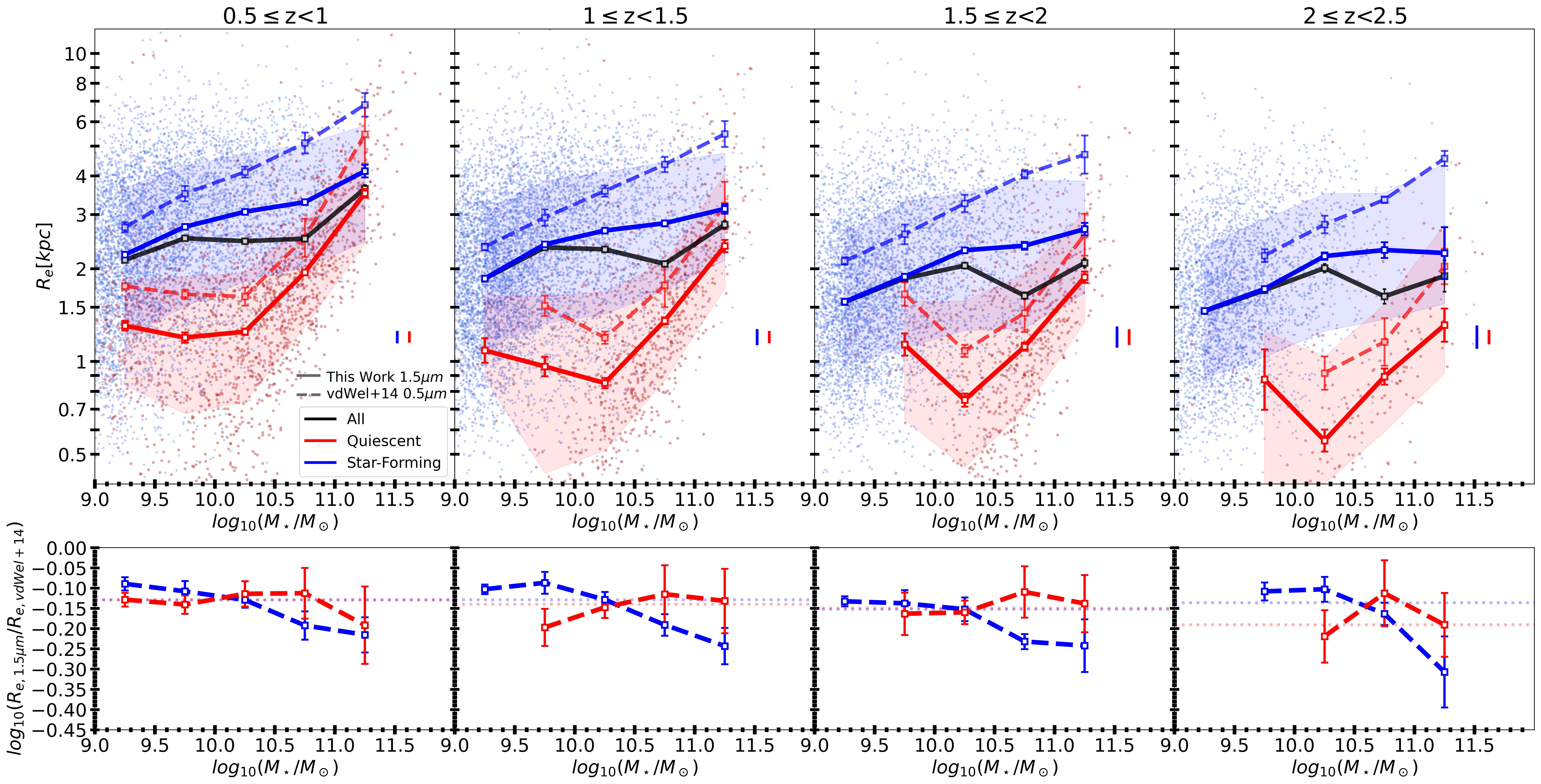}
    \caption{Size-Mass distribution at rest-frame 1.5$\mu{\text{m}}$ in four redshift bins. Galaxies are color-coded as quiescent (red) or star-forming (blue). Solid lines show the medians in mass bins of 0.5~dex for the two populations and the combined sample (black). Error bars represent the statistical uncertainty ($\sigma/\sqrt{N}$) while shaded areas highlight the 16-84 percentile intervals of the distributions. The median uncertainty on the rest-frame 1.5$\mu{\text{m}}$ size for star-forming and quiescent galaxies is shown in the bottom right corner of each panel.
    The 16-50-84 percentile values are reported in Table \ref{tab:med_sizes}.
    Dashed lines show the median relations in the rest-frame optical (0.5$\mu{\text{m}}$) from \cite{van-der-wel14} with shifted stellar mass values to account for systematic differences with the stellar masses used in this paper (see text and Appendix \ref{app-sec: Fast-LePhare} for details).
    Error bars {in the bottom right corner} represent the typical uncertainty of the individual size estimates.
    In the bottom panels, we show the offset between the median rest-frame near-IR and optical sizes. The overall offset for each redshift bin and galaxy type is indicated by the dotted horizontal lines. 
    }
    \label{fig:size_mass}
\end{figure*}

\begin{table*}
\caption{Median stellar mass ($M_{\star,\text{med}}$) and 16-50-84 percentiles of the effective radii (log$_{10}$($R_{\text{e}}$/{\text{kpc}})) presented in Figure \ref{fig:size_mass} respectively for the population as a whole (upper table), the star-forming population (central table) and the quiescent population (lower table) divided in mass and redshift bins. \label{tab:med_sizes}}
\begin{adjustbox}{center, pageouter, width=0.94\textwidth}\begin{tabular}{|c|cccc|cccc|cccc|cccc|}
    \hline
    \multicolumn{17}{|c|}{FULL SAMPLE} \\[1.8ex]
    \hline
     & \multicolumn{4}{c|}{$0.5\leq z <1.0$} & \multicolumn{4}{c|}{$1.0\leq z <1.5$}  & \multicolumn{4}{c|}{$1.5\leq z <2.0$} & \multicolumn{4}{c|}{$2.0\leq z <2.5$} \\
    \hline
    Mass Bin & $M_{\star,\text{med}}$ & \multicolumn{3}{c|}{log$_{10}$($R_{\text{e}}$)} & $M_{\star,\text{med}}$ & \multicolumn{3}{c|}{log$_{10}$($R_{\text{e}}$)} & $M_{\star,\text{med}}$ & \multicolumn{3}{c|}{log$_{10}$($R_{\text{e}}$)} & $M_{\star,\text{med}}$ & \multicolumn{3}{c|}{log$_{10}$($R_{\text{e}}$)} \\
     & 
     & 16\% & 50\% & 84\% &
     & 16\% & 50\% & 84\% &
     & 16\% & 50\% & 84\% & 
     & 16\% & 50\% & 84\% \\[0.4ex]
    \hline
    $[9.0, 9.5)$  & 9.24 & 0.07 & 0.33 & 0.56 & 9.25 & 0.04 & 0.27 & 0.49 &
                    9.27 & -0.01 & 0.19 & 0.42 &
                    9.27 & -0.05 & 0.16 & 0.37 \\[1.5ex]
    $[9.5, 10.0)$ & 9.72 & 0.11 & 0.40 & 0.62 & 9.72 & 0.11 & 0.37 & 0.58 &
                    9.70 & 0.03 & 0.27 & 0.51 &
                    9.68 & 0.01 & 0.23 & 0.46  \\[1.5ex]
    $[10.0, 10.5)$& 10.23 & 0.06 & 0.39 & 0.64 & 10.23 & 0.00 & 0.36 & 0.60 &
                    10.22 & -0.04 & 0.31 & 0.53 &
                    10.19 & -0.04 & 0.30 & 0.52 \\[1.5ex]
    $[10.5, 11.0)$& 10.71 & 0.19 & 0.40 & 0.62 & 10.72 & 0.04 & 0.32 & 0.58 &
                    10.71 & -0.05 & 0.21 & 0.51 &
                    10.69 & -0.12 & 0.21 & 0.51 \\[1.5ex]
    $[11.0, 11.5)$& 11.15 & 0.39 & 0.56 & 0.75 & 11.13 & 0.28 & 0.44 & 0.65 &
                    11.12 & 0.13 & 0.32 & 0.55 & 
                    11.16 & 0.03 & 0.28 & 0.54  \\[1.6ex]
    \toprule
    \multicolumn{17}{|c|}{STAR-FORMING} \\[1.8ex]
    \hline
     & \multicolumn{4}{c|}{$0.5\leq z <1.0$} & \multicolumn{4}{c|}{$1.0\leq z <1.5$}  & \multicolumn{4}{c|}{$1.5\leq z <2.0$} & \multicolumn{4}{c|}{$2.0\leq z <2.5$} \\
    \hline
    Mass Bin & $M_{\star,\text{med}}$ & \multicolumn{3}{c|}{log$_{10}$($R_{\text{e}}$)} & $M_{\star,\text{med}}$ & \multicolumn{3}{c|}{log$_{10}$($R_{\text{e}}$)} & $M_{\star,\text{med}}$ & \multicolumn{3}{c|}{log$_{10}$($R_{\text{e}}$)} & $M_{\star,\text{med}}$ & \multicolumn{3}{c|}{log$_{10}$($R_{\text{e}}$)} \\
     & 
     & 16\% & 50\% & 84\% &
     & 16\% & 50\% & 84\% &
     & 16\% & 50\% & 84\% & 
     & 16\% & 50\% & 84\% \\[0.4ex]
    \hline
    $[9.0, 9.5)$  & 9.23 & 0.08 & 0.35 & 0.56 & 9.25 & 0.04 & 0.27 & 0.49 &
                    9.27 & -0.01 & 0.20 & 0.42 &
                    9.27 & -0.05 & 0.17 & 0.37 \\[1.5ex]
    $[9.5, 10.0)$ & 9.71 & 0.18 & 0.44 & 0.64 & 9.71 & 0.12 & 0.38 & 0.59 &
                    9.70 & 0.04 & 0.28 & 0.52 &
                    9.68 & 0.01 & 0.23 & 0.46 \\[1.5ex]
    $[10.0, 10.5)$& 10.21 & 0.26 & 0.49 & 0.67 & 10.22 & 0.15 & 0.42 & 0.62 &
                    10.20 & 0.10 & 0.36 & 0.55 &
                    10.18 & 0.09 & 0.34 & 0.54 \\[1.5ex]
    $[10.5, 11.0)$& 10.68 & 0.31 & 0.52 & 0.70 & 10.70 & 0.23 & 0.44 & 0.65 &
                    10.69 & 0.11 & 0.37 & 0.58 &
                    10.65 & 0.14 & 0.36 & 0.55  \\[1.5ex]
    $[11.0, 11.5)$& 11.14 & 0.39 & 0.61 & 0.77 & 11.13 & 0.33 & 0.49 & 0.67 &
                    11.10 & 0.22 & 0.42 & 0.59 &
                    11.22 & 0.19 & 0.35 & 0.63 \\[1.6ex]
    \toprule
    \multicolumn{17}{|c|}{QUIESCENT} \\[1.8ex]
    \hline
     & \multicolumn{4}{c|}{$0.5\leq z <1.0$} & \multicolumn{4}{c|}{$1.0\leq z <1.5$}  & \multicolumn{4}{c|}{$1.5\leq z <2.0$} & \multicolumn{4}{c|}{$2.0\leq z <2.5$} \\
    \hline
    Mass Bin & $M_{\star,\text{med}}$ & \multicolumn{3}{c|}{log$_{10}$($R_{\text{e}}$)} & $M_{\star,\text{med}}$ & \multicolumn{3}{c|}{log$_{10}$($R_{\text{e}}$)} & $M_{\star,\text{med}}$ & \multicolumn{3}{c|}{log$_{10}$($R_{\text{e}}$)} & $M_{\star,\text{med}}$ & \multicolumn{3}{c|}{log$_{10}$($R_{\text{e}}$)} \\
     & 
     & 16\% & 50\% & 84\% &
     & 16\% & 50\% & 84\% &
     & 16\% & 50\% & 84\% & 
     & 16\% & 50\% & 84\% \\[0.4ex]
    \hline
    $[9.0, 9.5)$  & 9.26 & -0.08 & 0.12 & 0.26 & 9.36 & -0.05 & 0.03 & 0.22 &
                    -- & -- & -- & -- &
                    -- & -- & -- & -- \\[1.5ex]
    $[9.5, 10.0)$ & 9.78 & -0.17 & 0.08 & 0.27 & 9.80 & -0.37 & -0.02 & 0.22 &
                    9.80 & -0.19 & 0.05 & 0.19 &
                    9.74 & -0.47 & -0.06 & 0.10  \\[1.5ex]
    $[10.0, 10.5)$& 10.30 & -0.13 & 0.10 & 0.29 & 10.29 & -0.29 & -0.07 & 0.22 &
                    10.32 & -0.35 & -0.13 & 0.19 &
                    10.32 & -0.39 & -0.26 & 0.02 \\[1.5ex]
    $[10.5, 11.0)$& 10.73 & 0.11 & 0.29 & 0.45 & 10.76 & -0.06 & 0.13 & 0.33 &
                    10.76 & -0.16 & 0.05 & 0.29 &
                    10.77 & -0.23 & -0.04 & 0.19 \\[1.5ex]
    $[11.0, 11.5)$& 11.16 & 0.38 & 0.55 & 0.73 & 11.12 & 0.24 & 0.37 & 0.61 &
                    11.14 & 0.12 & 0.27 & 0.48 &
                    11.13 & -0.05 & 0.12 & 0.44 \\[1.6ex]
    \hline
\end{tabular}
\end{adjustbox}
\end{table*}

In Figure \ref{fig:size_mass} we present the rest-frame near-IR size-stellar mass distribution 
and its evolution from $z=0.5$ to $2.5$. 
Stellar masses from the COSMOS2020 catalog are adjusted to account for the difference between the total amount $K_s-$band light that entered the stellar mass estimate and the total amount of F277W light in our fitted profiles; see Appendix \ref{app-sec: Mass-color} for details.

$R_{\text{e},1.5\mu{\text{m}}}$ does not strongly correlate with stellar mass: at $z=1-1.5$ both $M_\star=10^{11}~{\text{M}}_\odot$ and $M_\star=10^{9}~{\text{M}}_\odot$~galaxies have typical values of $R_{\text{e},1.5\mu{\text{m}}}\approx 2$~{\text{kpc}}, despite 2 orders of magnitude difference in stellar mass. This relative flatness of the size-mass relation is seen across all 6~Gyr of cosmic time spanned by our sample (black lines in Fig.~\ref{fig:size_mass}).  At the same time, the variety in sizes is large. The full range spans at least 1.5 orders of magnitude (at fixed stellar mass) and the 1$\sigma$ scatter is 0.3-0.4~dex, slightly depending on mass and redshift. 

As is well known, much of this scatter correlates with star-formation activity.  After separating into star-forming and quiescent subsamples, the familiar patterns appear. Quiescent galaxies are consistently smaller than their star-forming counterparts at all redshifts \citep[e.g.,][]{franx08, van-der-wel14}. The maximum difference in size is seen at $M_\star\approx10^{10-10.5}~{\text{M}}_\odot$, reaching a factor $\approx 3$ at $z>1$. The size-mass relation for the quiescent population is steeply increasing for $M_\star>10^{10}~{\text{M}}_\odot$, and flat at lower masses. The size-mass relation for star-forming galaxies is flatter but also monotonically increasing. The flatness in the size-mass relation for the full population is due to the joint effect of differences in the mass functions and size distributions of quiescent and star-forming galaxies.

At all redshifts and stellar masses, we see that $R_{\text{e},1.5\mu{\text{m}}}$ is, on average, smaller than the $R_{\text{e},0.5\mu{\text{m}}}$ rest-frame optical ($0.5~\mu{\text{m}}$) size measurements from \cite{van-der-wel14}. The comparison with \cite{van-der-wel14} is made after correcting for a slight difference between the adopted cosmologies and, more importantly, differences in the stellar mass estimates, which are substantial and redshift-dependent. These corrections are outlined in Appendix \ref{app-sec: Fast-LePhare}. 
The dashed lines in Figure \ref{fig:size_mass} represent the median values of the reconstructed \cite{van-der-wel14} size-mass distribution.

The offset between the optical and near-IR $R_{\text{e}} $ is generally $-0.10$ to $-0.15$~dex, mostly independent of redshift, galaxy type and mass: galaxies are 25-40\% smaller at 1.5$~\mu{\text{m}}$ than at 0.5$~\mu{\text{m}}$. The one discernible deviation from this baseline offset is that high-mass ($M_\star>10^{10.5}~{\text{M}}_\odot$) star-forming galaxies show larger offsets, especially at the highest redshift, reaching up to a factor 2.

\cite{suess22} were the first to directly compare rest-frame optical and near-IR sizes of galaxies in this redshift range, using F150W and F444W imaging from CEERS \citep{Finkelstein23}. This corresponds to rest-frame 0.5$\mu{\text{m}}$ and 1.5$\mu{\text{m}}$ at $z\approx 2$. They found a median difference of $\approx -0.10$~dex, and a negative trend with mass, in fair agreement with our findings. One notable difference, at first sight, is the lack of an offset in their F150W-to-F444W size comparison for low-mass star-forming galaxies ($9<\log(M_\star/{\text{M}}_\odot)<10$). Upon closer inspection, the difference is minimal: their offset of -0.03~dex becomes -0.05~dex when comparing their F444W sizes to $R_{\text{e},0.5\mu m}$ from \citet{van-der-wel14}, and a direct comparison of overlapping galaxies in our sample also produces $R_{\text{e},F444W}/R_{\text{e},0.5\mu m}=-0.05$~dex when using \textsc{WebbPSF} and $-$0.08~dex when using our default stacked PSF. It is safe to conclude that these lower-mass star-forming galaxies show some offset in $R_{\text{e},1.5\mu m}/R_{\text{e},0.5\mu m}$, but whether this is closer to $-0.05$~dex or $-0.10$~dex remains to be confirmed.

Very recently, \cite{ji24} showed how the sizes of a smaller sample of 151 quiescent galaxies at $z>1$ depend on wavelength -- from rest-frame 0.3$~\mu{\text{m}}$ to 1$~\mu{\text{m}}$ -- based on data from JADES \citep[PIs Rieke
and Lutzgendorf;][]{eisenstein23}.
They find typical systematic differences of $-0.08$~dex when comparing $R_{\text{e},1\mu m}$ and $R_{\text{e},0.5\mu{\text{m}}}$, and little variation in this offset with mass or redshift. These results are consistent with the results presented in this paper considering that we extend to longer wavelengths. 

{\cite{van-der-wel24} presented a comparison at redshifts $0.5<z<1.5$ between stellar half-mass radii and rest-frame near-IR half-light radii from JWST/CEERS in \cite{martorano23}. They found excellent agreement, with differences below 0.03~dex at rest-frame 2$\mu$m, suggesting that the offsets presented in this paper largely account for the differences between optical sizes and stellar mass-weighted sizes.}

\subsection{Redshift evolution}\label{sec:z-evo}

\begin{figure*}
    \centering
    \includegraphics[scale=0.4]{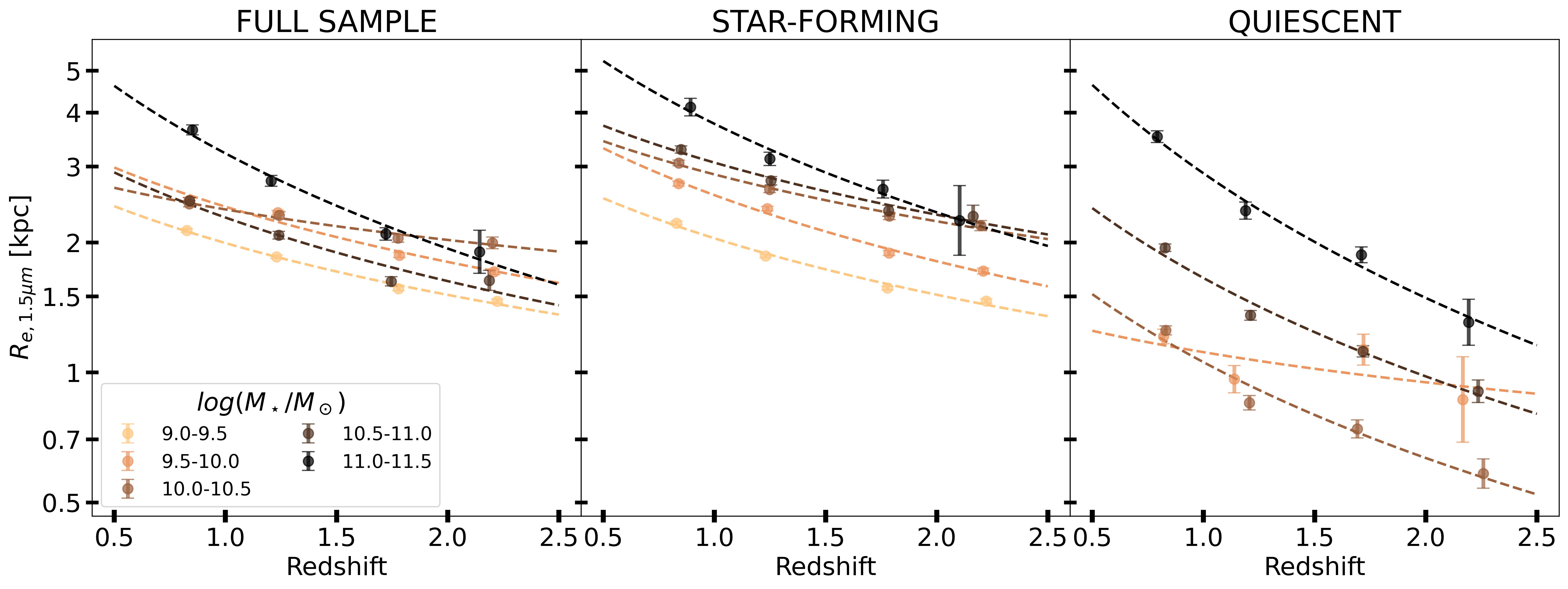}
    \caption{Redshift evolution of median sizes for the whole sample (left panel), star-forming galaxies (central panel), and quiescent galaxies (right panel) in five stellar mass bins. Error bars show the statistical uncertainty expressed as $\sigma/\sqrt{N}$ while dotted lines show fit to equation \ref{Eq:Fit_eq}. Fit coefficients are reported in Table \ref{tab:fit_med_size_BOTH}.}
    \label{fig:size_evo}
\end{figure*}

In Figure \ref{fig:size_evo} we explicitly show the evolution $R_{\text{e},1.5\mu{\text{m}}}$ as a function of redshift and stellar mass. 
We parameterize the evolution as a function of redshift and Hubble parameter:
\begin{equation} \label{Eq:Fit_eq}
    \log_{10}(R_{\text{e}} ) = \alpha_1+\beta_1 \log_{10}(1+z)
\end{equation}
\begin{equation} \label{Eq:Fit_eq_h(z)}
    \log_{10}(R_{\text{e}} ) = \alpha_2+\beta_2 \log_{10}\left(\frac{H(z)}{H_0}\right)
\end{equation}

Fits are repeated 10.000 times, with Gaussian sampling the median sizes from their uncertainty distributions. The median of the coefficients retrieved from the multiple fits and their 16-84 percentiles are reported in Table \ref{tab:fit_med_size_BOTH}. Figure \ref{fig:size_evo} shows only the redshift parameterization.

The sizes of star-forming galaxies with $M_\star < 10^{11}~{\text{M}}_\odot$ evolve as $R_{\text{e},1.5\mu{\text{m}}} \propto (1+z)^{\approx -0.7}$, with little dependence on mass. Only higher-mass galaxies show substantially faster evolution: $R_{\text{e},1.5\mu{\text{m}}} \propto (1+z)^{\approx -1.2}$, comparable with the evolution of quiescent galaxies. Quiescent galaxies generally evolve faster in size: $R_{\text{e},1.5\mu{\text{m}}} \propto (1+z)^{\approx -1.3}$ for $M_\star > 10^{10}~{\text{M}}_\odot$. Lower-mass quiescent galaxies evolve slower ($R_{\text{e},1.5\mu{\text{m}}} \propto (1+z)^{\approx -0.4}$). Note that the combined population (SF$+$Q) evolves in size slower than star-forming alone, which is related to the increasing fraction of quiescent galaxies with cosmic time.

We find consistent results when adopting the $R_{\text{e}} \propto H(z)^{\beta_2}$ parameterization with $R_{\text{e},1.5\mu{\text{m}}}\propto H(z)^{-0.6}$ and $R_{\text{e},1.5\mu{\text{m}}}\propto H(z)^{-1.0}$, respectively, for the star-forming and quiescent population. The offset between $\beta_2$ and $\beta_1$ reflects the different evolution with cosmic time of the parameters $(1+z)$ and $H(z)$, with the latter evolving slower at late times.

\begin{table*}
\centering
\caption{Fit results of the redshift evolution of median sizes as parameterized in Equations \ref{Eq:Fit_eq} and \ref{Eq:Fit_eq_h(z)}. Values in the table represent the median and 16-84 percentiles of $10\,000$ fits performed on the median sizes in redshift bins presented in Table \ref{tab:med_sizes} sampled from their Gaussian distribution with $\sigma$ equal to the statistical uncertainty.}
\label{tab:fit_med_size_BOTH}
\begin{adjustbox}{center, textareainner, width=0.93\textwidth}\begin{tabular}{|c|cc|cc|cc|}
    \hline
    \multicolumn{7}{|c|}{redshift parametrization (Eq.\ref{Eq:Fit_eq})}\\
    \hline
    & \multicolumn{2}{c|}{FULL SAMPLE} & \multicolumn{2}{c|}{STAR-FORMING} & \multicolumn{2}{c|}{QUIESCENT}\\
        \hline
        Mass Bins & $\alpha_1$ & $\beta_1$ & $\alpha_1$ & $\beta_1$ & $\alpha_1$ & $\beta_1$ \\[0.4ex]
        \hline
        $[9.0, 9.5)$ & 
        $0.50^{+0.01}_{-0.01}$ & $-0.68^{+0.03}_{-0.03}$ & $0.53^{+0.01}_{-0.01}$ & $-0.74^{+0.02}_{-0.03}$ & -- & --\\[2.1ex]$[9.5, 10.0)$ &
        $0.60^{+0.01}_{-0.01}$ & $-0.73^{+0.03}_{-0.03}$ & $0.67^{+0.01}_{-0.01}$ & $-0.87^{+0.03}_{-0.03}$ & $0.17^{+0.12}_{-0.11}$ & $-0.40^{+0.35}_{-0.37}$\\[2.1ex]$[10.0, 10.5)$ & 
        $0.50^{+0.02}_{-0.02}$ & $-0.40^{+0.06}_{-0.06}$ & $0.64^{+0.02}_{-0.02}$ & $-0.62^{+0.05}_{-0.05}$ & $0.40^{+0.04}_{-0.04}$ & $-1.26^{+0.13}_{-0.13}$\\[2.1ex] $[10.5, 11.0)$ & 
        $0.61^{+0.03}_{-0.03}$ & $-0.84^{+0.09}_{-0.09}$ & $0.69^{+0.03}_{-0.03}$ & $-0.69^{+0.10}_{-0.09}$ & $0.61^{+0.03}_{-0.03}$ & $-1.29^{+0.10}_{-0.10}$\\[2.1ex] $[11.0, 11.5)$ & 
        $0.88^{+0.06}_{-0.06}$ & $-1.25^{+0.19}_{-0.19}$ & $0.93^{+0.10}_{-0.11}$ & $-1.16^{+0.31}_{-0.31}$ & $0.95^{+0.06}_{-0.06}$ & $-1.64^{+0.19}_{-0.19}$\\[2.1ex] \toprule
    \multicolumn{7}{|c|}{cosmological parametrization (Eq.\ref{Eq:Fit_eq_h(z)})}\\
    \hline
    & \multicolumn{2}{c|}{FULL SAMPLE} & \multicolumn{2}{c|}{STAR-FORMING} & \multicolumn{2}{c|}{QUIESCENT}\\
        \hline
        Mass Bins & $\alpha_2$ & $\beta_2$ & $\alpha_2$ & $\beta_2$ & $\alpha_2$ & $\beta_2$ \\[0.4ex]
        \hline
        $[9.0, 9.5)$ & 
        $0.43^{+0.01}_{-0.01}$ & $-0.53^{+0.02}_{-0.02}$ & $0.45^{+0.01}_{-0.01}$ & $-0.58^{+0.02}_{-0.02}$ & -- & --\\[2.1ex] $[9.5, 10.0)$ &
        $0.53^{+0.01}_{-0.01}$ & $-0.57^{+0.03}_{-0.03}$ & $0.58^{+0.01}_{-0.01}$ & $-0.68^{+0.03}_{-0.03}$ & $0.13^{+0.08}_{-0.08}$ & $-0.32^{+0.29}_{-0.28}$\\[2.1ex] $[10.0, 10.5)$ & 
        $0.46^{+0.01}_{-0.01}$ & $-0.31^{+0.05}_{-0.04}$ & $0.58^{+0.01}_{-0.01}$ & $-0.48^{+0.04}_{-0.04}$ & $0.26^{+0.03}_{-0.03}$ & $-0.98^{+0.11}_{-0.10}$\\[2.1ex] $[10.5, 11.0)$ & 
        $0.52^{+0.02}_{-0.02}$ & $-0.65^{+0.07}_{-0.07}$ & $0.62^{+0.02}_{-0.02}$ & $-0.54^{+0.08}_{-0.08}$ & $0.47^{+0.02}_{-0.02}$ & $-1.01^{+0.08}_{-0.08}$\\[2.1ex] $[11.0, 11.5)$ & 
        $0.75^{+0.04}_{-0.04}$ & $-0.98^{+0.15}_{-0.15}$ & $0.80^{+0.07}_{-0.07}$ & $-0.91^{+0.25}_{-0.25}$ & $0.78^{+0.04}_{-0.04}$ & $-1.29^{+0.15}_{-0.16}$\\ \hline
    \end{tabular}
\end{adjustbox}
\end{table*}

\begin{figure*}
    \centering
    \includegraphics[scale=0.5]{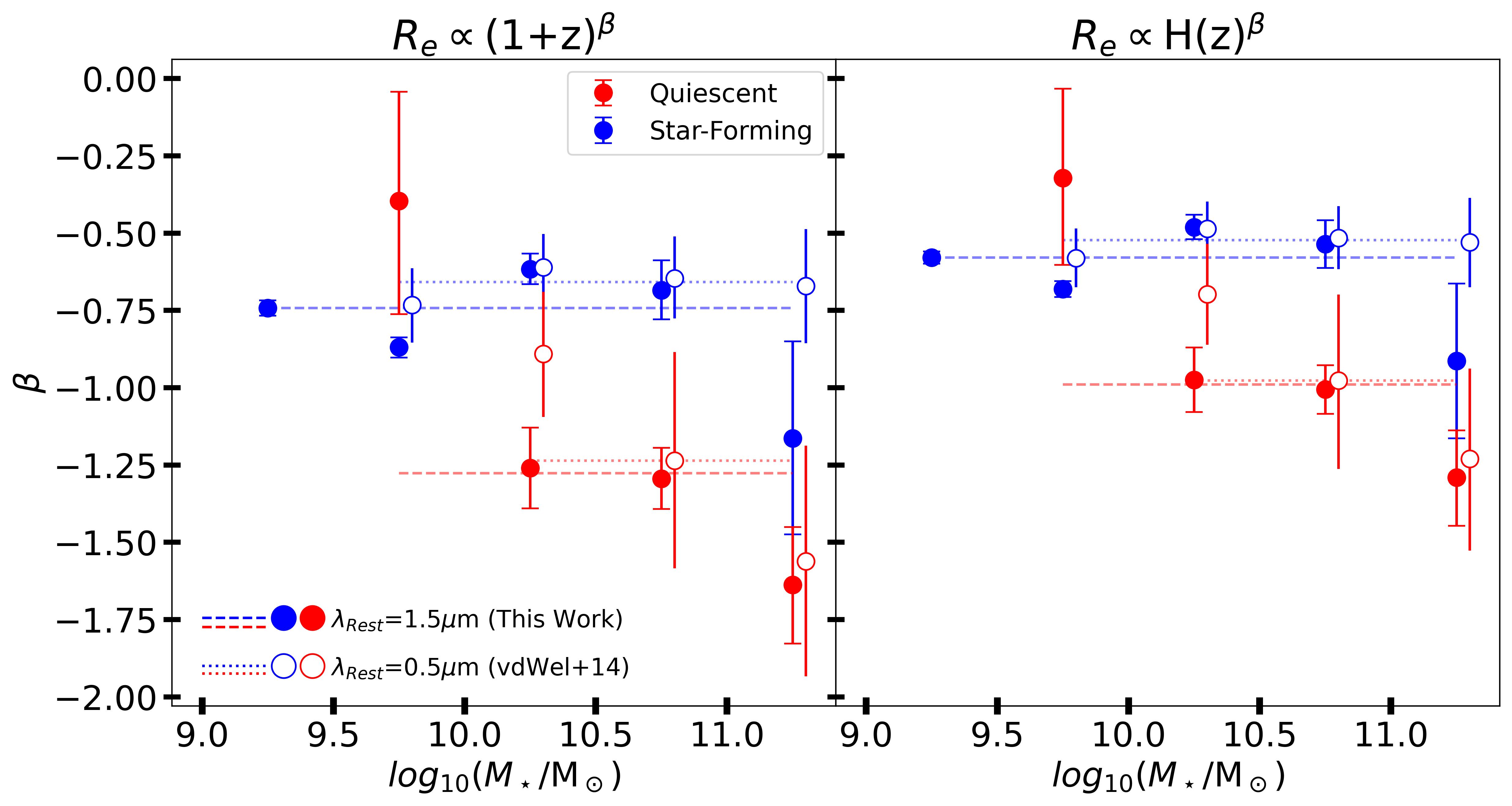}
    \caption{Size evolution coefficients $\beta$ obtained with the redshift (left panel) and Hubble parameterizations (right panel). We compare our results at rest-frame 1.5~$\mu{\text{m}}$ (filled circles) with those at 0.5~$\mu{\text{m}}$ recomputed from \cite{van-der-wel14} data (open circles). In blue, we show coefficients relative to the star-forming galaxies, while in red, we show those retrieved for the quiescent population. Dashed horizontal lines show the median of the $\beta$ coefficients retrieved in this work, while dotted lines are used for the median of \cite{van-der-wel14} $\beta$ coefficients. For both cases, we show just the coefficients retrieved fitting median sizes in all four redshift bins investigated. \cite{van-der-wel14} values are shifted in mass by 0.05~dex to enhance readability.}
    \label{fig:param_comparison}
\end{figure*}

In Figure \ref{fig:param_comparison} we compare the pace of size evolution with redshift at 0.5$~\mu{\text{m}}$ and 1.5$~\mu{\text{m}}$. We use the revised mass estimates for the \cite{van-der-wel14} sample as described above and fit the same parameterized functions to their $R_{\text{e},0.5\mu{\text{m}}}$ sizes. There is remarkable agreement, implying the absence of strong wavelength dependencies in the pace of size evolution. This reinforces the recent results from JADES for a smaller sample of quiescent galaxies \citep{ji24}. 
Similarly, we find consistent results for $R_{\text{e}} \propto H(z)^{\beta_2}$ at rest-frame 0.5$~\mu{\text{m}}$ and 1.5$~\mu{\text{m}}$.

That said, the highest-mass star-forming galaxies and quiescent galaxies with stellar masses $10<~\log(M_\star/{\text{M}}_\odot)<10.5$ tentatively show faster evolution at 1.5~$\mu{\text{m}}$ than at 0.5~$\mu{\text{m}}$. 
These notable deviations are discussed further in \S\ref{sec:pace}.

\section{DISCUSSION} \label{sec:discussion}

\subsection{Flattening of the size-mass relation}\label{sec:flattening}
As shown in Figure \ref{fig:size_mass} both quiescent and star-forming galaxies are systematically smaller at rest-frame 1.5$~\mu{\text{m}}$ than at 0.5$~\mu{\text{m}}$ by 0.14$\pm0.03$~dex when averaging over all masses $M_\star > 10^9~{\text{M}}_\odot$, and in a manner that is approximately independent of redshift. The offset is also independent of star-formation activity; even though color gradients are thought to arise through different effects -- attenuation is important for star-forming galaxies \citep[e.g.,][]{Miller2022}, while for quiescent galaxies, the difference arises due to metallicity/age gradients in the stellar populations -- the end effect is, on average, similar. Overall, the same patterns appear for the size-mass relation in the rest-frame near-IR as seen in the rest-frame optical, with a steep relation for massive quiescent galaxies and a flatter relation for star-forming galaxies.

The only relevant and statistically significant correlation is the one between $M_\star$ and $R_{\rm{e,1.5\mu{\text{m}}}}/R_{\rm{e,0.5\mu{\text{m}}}}$ for star-forming galaxies (see Fig. \ref{fig:size_mass}). Low-mass galaxies ($M_\star \approx 10^9~{\text{M}}_\odot$) are $\sim$0.1~dex smaller at 1.5$~\mu{\text{m}}$, while high-mass galaxies ($M_\star > 10^{11}~{\text{M}}_\odot$) are $\sim$0.25~dex smaller: the size-mass relation flattens. In addition, while at $z>1.5$ this mass dependence is manifest only for $M_\star > 10^{10.5}~{\text{M}}_\odot$, at $z<1$ it is seen across the entire mass range. There is a 0.1-0.15~dex baseline offset between $R_{\rm{e,1.5\mu{\text{m}}}}$ and $R_{\rm{e,0.5\mu{\text{m}}}}$ for all galaxies, regardless of redshift, mass, and galaxy type, with an additional offset for star-forming galaxies above a redshift-dependent stellar mass limit. These results echo those from \citet{lange15}, who identified a similar pattern for present-day spiral galaxies.

We note that when we repeat the entire analysis with \textsc{WebbPSF}~PSFs instead of the stacked stars, we find sizes that are systematically 0.03~dex($\pm$0.02~dex) larger. This systematic effect has no significant impact on any of our results, which  {ensures that our results are robust against systematic errors in the adopted PSFs.}

The flattening of the size-mass relation at rest-frame 1.5$~\mu{\text{m}}$ for massive star-forming galaxies and the almost systematic shift between optical and near-IR sizes that affects all the galaxies, independently of their classification, are likely caused by a combination of different phenomena.
The systematic shift between sizes at 0.5$~\mu{\text{m}}$ and 1.5$~\mu{\text{m}}$ implies that rest-frame near-IR profiles are more compact than optical profiles \citep{dutton11, wuyts12, guo12, szomoru13, cibinel15, suess22}, which presumably implies that stellar mass profiles are more centrally concentrated than optical light profiles.
The color and $M_\star/L$ gradients reflect a combination of dust concentration in the central region of the galaxy \citep{de-jong96a, bell01, nelson16, Miller2022, zhang23, lebail23} and gradients in stellar populations \citep[i.e.][]{bezanson09, bezanson11, nelson19, tadaki20, deugenio20}. Since dust content increases with galaxy mass \citep{bell03b, garn10, zahid13, alvarez-marquez16, whitaker17, gomez-guijarro23}, a mass-dependent effect of attenuation on galaxy size estimates is certainly plausible. Furthermore, edge-on galaxies are reddened \citep[e.g.,][]{patel13}, with bright centers potentially obscured at shorter wavelengths. Visual inspection of such sources confirms that this incidentally occurs, but this effect is not dominant. More massive galaxies also have stronger stellar population gradients. The separation of the various effects must be assessed via simulations \citep[i.e.][]{baes24b} and by spatially resolved SED modeling at high redshift across the full wavelength range provided by HST and JWST \citep[see, e.g.,][]{abdurrouf23}. Another fruitful approach would be to analyze color gradients and $M_\star/L$ gradients at low redshift \citep[i.e.][]{ge21} where spatially resolved, high-SNR spectra are available.

\subsection{Size evolution with redshift}
\label{sec:pace}

The pace of size evolution with redshift at 1.5$\mu{\text{m}}$ and 0.5$\mu{\text{m}}$ is very similar (see Figures \ref{fig:size_mass} and \ref{fig:param_comparison}), with a moderate pace of evolution, $\propto (1+z)^{-0.7}$, for star-forming galaxies and a faster pace, $\propto (1+z)^{-1.3}$, for massive quiescent galaxies ($M_\star > 10^{10}~{\text{M}}_\odot$). For lower-mass quiescent galaxies, the pace of evolution is slower, similar to that for star-forming galaxies.  The fact that the choice of wavelength has a negligible effect on the observed pace of size evolution strongly suggests that attenuation and stellar population gradients have no substantial impact on the interpretation of the observed pace of size evolution. In other words, whereas light-weighted sizes underestimate the true, stellar mass-weighted sizes of galaxies, this effect shows no discernible redshift dependence in the range $0.5<z<2.5$. We can conclude with good confidence that the observed pace of size evolution accurately reflects the pace of evolution in mass-weighted sizes. The one remaining caveat is that biases that we cannot control for may exist that are independent of wavelength but dependent on redshift. Like the UV and optical, the rest-frame near-IR is not a direct stellar mass tracer; variations in age, metallicity, and, in severe cases, attenuation lead to variations in $M/L$ at all wavelengths. 

Two deviations from this narrative are worth mentioning. First, the most massive star-forming galaxies evolve in size faster than less massive star-forming galaxies, but only at 1.5$~\mu{\text{m}}$, not at 0.5$~\mu{\text{m}}$. This is directly connected to the flattening of the size-mass relation that is most pronounced at higher redshifts (see \S\ref{sec:flattening} for a discussion).
Second, quiescent galaxies with stellar mass $10^{10}<\log(M_\star/{\text{M}}_\odot)<10^{10.5}$ show fast evolution in size at 1.5$\mu{\text{m}}$ but slower evolution at 0.5$\mu{\text{m}}$. The median value is $R_{\text{e},1.5\mu{\text{m}}}\approx 0.6$~{\text{kpc}} at $z=2$, below the HST resolution limit, but the tentative evidence based on a small sample is that even at JWST resolution, this difference remains \citep{ji24}.  These very compact galaxies either have somewhat attenuated young centers, in line with stellar population gradients seen in compact post-starburst galaxies $z\approx 1$ \citep{deugenio20}, or low-level residual star formation in their outer parts. Larger samples along with detailed color gradient measurements are needed to confirm this result.

In general, our results match those from \citet{van-der-wel24}, who showed very similar behavior in $R_{\text{e},M\star}/R_{\rm{e,0.5\mu{\text{m}}}}$ as we now see for $R_{\rm{e,1.5\mu{\text{m}}}}/R_{\rm{e,0.5\mu{\text{m}}}}$: similar pace of evolution along with a flattening of the size-mass relation for star-forming galaxies. It is of particular interest to note that the early evidence is that the near-IR sizes of quiescent galaxies are very similar to their rest-frame optical sizes, modulo an offset that is mostly independent of redshift. This contradicts the conclusion from 
\cite{suess19a, suess19b} and \citet{miller23} that stellar mass-weighted sizes for quiescent galaxies evolve slowly (or not at all) at $z>1$. We refer to the Appendix of  \citet{van-der-wel24} for further discussion.

In conclusion, the wavelength dependence of galaxy sizes and the consequent color gradients do not strongly distort the size evolution seen in the rest-frame optical. The observed pace of size evolution of quiescent galaxies, regardless of wavelength, reflects a combination of a progenitor bias -- newly quenched galaxies have different properties than previously quenched galaxies \citep{van-dokkum01a} -- that contributes to the pace of size evolution observed for samples of galaxies \citep{van-der-wel09, carollo16, wu18b, ji22} on top of the evolution of individual galaxies \citep{van-der-wel09, bezanson09}.

\section{OUTLOOK} \label{sec:conclusion}
In this work, we analyze the rest-frame 1.5~$\mu{\text{m}}$ light profiles of {$\sim26\,000$} galaxies in the COSMOS field with NIRCam imaging in the F277W and F444W filters from the COSMOS-WEB and PRIMER surveys. The sample is drawn from the preexisting COSMOS2020 catalog \citep{weaver22} and not NIRCam-selected. The main caveat in our work is therefore the mismatch between the source catalog and what is seen in the NIRCam imaging. A more extensive analysis that starts with NIRCam selection of sources and subsequent multiwavelength photometry and estimation of the key parameters (redshifts, stellar masses, etc.) and light profile fits is a logical next step. This will allow one to not only make a more consistent analysis of the size - stellar mass distribution and its evolution, but also extend it to higher redshifts. Despite this, we assert that the conclusions we draw in this work will likely not change once these improvements are made.

Furthermore, until now, the evolution of the rest-frame near-IR galaxy sizes has been treated as indicative of the evolution of stellar mass-weighted sizes, implicitly assuming that near-IR light profiles correspond with stellar mass profiles. In order to take full advantage of the wide wavelength range of spatially resolved light profiles provided by NIRCam for many thousands of galaxies across most of cosmic time, a rigorous analysis of the UV-to-near-IR color gradients is required. This is less straightforward than it may seem, as our knowledge of near-IR SEDs is poorly constrained, especially at large lookback times. For example, \citet{van-der-wel21} showed that the near-IR colors ($J-H$ and $H-K$) of $z\approx 1$ galaxies do not match the template colors of the most widely used stellar population models. The highly accurate and precise NIRCam photometry will produce SEDs with small systematic uncertainties, which will then allow us to assess which stellar population models and underlying isochrones and stellar template libraries match the data well. Only then are we in a good position to interpret the observed color gradients and reconstruct stellar mass profiles with high accuracy.

\begin{acknowledgments}
M.M. acknowledges the financial support of the Flemish Fund for Scientific Research (FWO-Vlaanderen), research project G030319N.
The data products presented herein were retrieved from the Dawn JWST Archive (DJA). DJA is an initiative of the Cosmic Dawn Center, which is funded by the Danish National Research Foundation under grant No. 140.
\end{acknowledgments}

\software{  astropy \citep{Astropy_v0.2, Astropy_v2, Astropy_v5},  
            \textsc{GalfitM} \citep{haussler13, vika13},
            \textsc{SEP} \citep{bertin96, barbary16} 
          }

\appendix

\section{Stellar mass corrections}

\subsection{Stellar Mass correction}\label{app-sec: Mass-color}
The stellar masses from the COSMOS2020 catalog are normalized by the measured total amount of $K_s$ band light, while our $R_{\text{e}} $ estimates are based on S\'ersic profile fits that estimate the total amount of light seen with NIRCam. To correct the stellar mass estimates for this difference, we add the difference between the F277W \textsc{GalfitM} flux and the $K_s$ total flux to the original stellar mass while taking into account the $K_s-F277W$ color:
\begin{equation}
    M_{\star,\text{corr}} = M_{\star,\text{LePhare}} + 0.4*(K_s - F277W - \langle K_s - F277W\rangle _{|J-K_s})
\end{equation}
Here $J$, $K_s$ and $F277W$ are the magnitudes of each galaxy, respectively, in the ULTRA-VISTA $J$ and $K_s$ bands and in the JWST/NIRCam filter F277W, with the latter taken from the \textsc{GalfitM} profile fit as outlined in Section \ref{sec:JWSTfit} while the former are retrieved from the COSMOS2020 catalog at an aperture of \SI{2}{\arcsecond} and corrected to be total magnitudes. $\langle K_s - F277W\rangle_{|J-K_s}$ is the median $K_s - F277W$ at fixed $J-K_s$, which produces the (on average) correct $K_s - F277W$ color correction for galaxies with a given $J-K_s$ color. This color is computed as the median $K_s - F277W$ in 0.25 mag wide bins of $J-K_s$, linearly interpolating in between the bin centers.

The use of $\langle K_s - F277W\rangle_{|J-K_s}$ instead of the global median of $K_s - F277W$ is motivated by the correlation between $K_s - F277W$ and $J-K_s$ that is negative for $J-K_s<0.4$, positive for $0.4<J-K_s<1$ and flat for redder colors inducing up to 0.3mag variations in the median $K_s - F277W$.

The resulting median correction is $M_{\star,\text{corr}} - M_{\star,\text{LePhare}}=0$, with a standard deviation of 0.12 dex. We note that these corrections cause slight shifts in the mass-size distribution but do not affect any of our conclusions. 

\subsection{FAST - LePhare matching}\label{app-sec: Fast-LePhare}
\begin{figure*}
    \centering
    \includegraphics[scale=0.303]{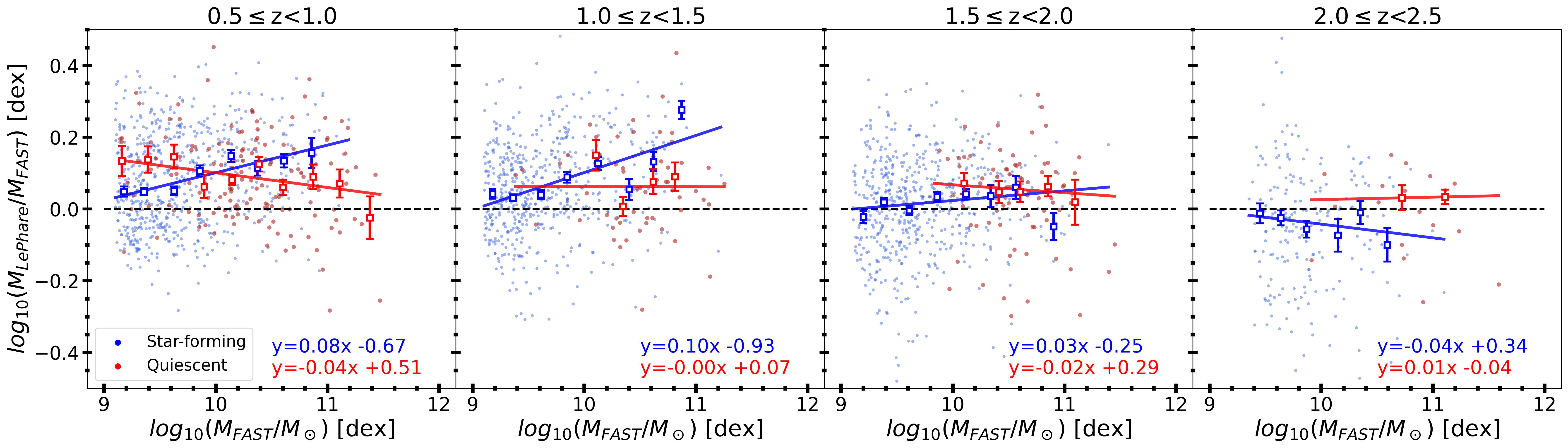}
    \caption{Stellar mass comparison between values used in \cite{van-der-wel14} and retrieved with \textsc{FAST} and those retrieved with \textsc{LePhare} on the COSMOS2020 photometry in \cite{weaver22} for cross-matched galaxies in the PRIMER-COSMOS field divided into four redshift bins. 
    Red and blue colors are used, respectively, for the quiescent and star-forming populations. Squares show the median in bins of 0.25dex with error bars representing the statistical uncertainty ($\sigma/\sqrt{N}$) while solid lines are the linear regression performed to the medians and used to match \textsc{FAST} to \textsc{LePhare} stellar masses. In the bottom right corner are shown the regression fit expressions.}
    \label{fig:M_comparison}
\end{figure*}

Stellar masses adopted in \cite{van-der-wel14} were retrieved with the package \textsc{FAST} \citep{kriek09b}. Cross-matching the sample used in this work and in \cite{van-der-wel14} which overlaps in the PRIMER-COSMOS field, we find $2\,267$ matches within \SI{0.4}{\arcsecond}. Comparing \textsc{LePhare} and \textsc{FAST} masses, we retrieved a substantial and redshift-dependent difference that we had to correct for a proper comparison of the samples.
We performed a simple linear regression fit to the mass difference for the cross-matched sample as a function of the \cite{van-der-wel14} $M_\star$ in the four redshift bins separately for quiescent and star-forming galaxies. Fit results are used to shift the $M_\star$ estimates of all the galaxies in the \cite{van-der-wel14} sample by the appropriate amount and create the size-mass distribution that can be directly compared to the current sample.

In Figure \ref{fig:M_comparison} we present the $M_{\star,\text{LePhare}}-M_{\star,\text{FAST}}$ difference as a function of $M_{\star,\text{FAST}}$ together with the results of the linear regression adopted for the correction.

To account for relative uncertainties in the two stellar mass estimates, reflected by a residual 1$\sigma$ scatter of 0.13 dex, we repeat $10\,000$ times the median size estimate in each mass and redshift bin, each time perturbing the adjusted \cite{van-der-wel14} $M_\star$ value by a random value drawn from a Gaussian with $\sigma=0.13$~dex.

\bibliography{mypapers.bib}{}
\bibliographystyle{aasjournal}

\end{document}